\documentclass[preprint,notoc]{JHEP3}
\usepackage{epsfig}

\def\dsl{\partial \llap/}
\def\sr2{\sqrt{2}}

\def\to{\rightarrow}

\def\bi{\begin{itemize}}
\def\ei{\end{itemize}}

\def\c1p{C1^\prime}
\def\ta{\tilde a}

\def\tu{\tilde u}

\def\ta{\tilde a}

\def\tst{\tilde t}

\def\tg{\tilde g}

\def\tq{\tilde q}

\def\tz{\widetilde Z}

\def\alt{\lesssim}
\def\agt{\gtrsim}
\def\be{\begin{equation}}
\def\ee{\end{equation}}
\def\bea{\begin{eqnarray}}
\def\eea{\end{eqnarray}}

\def\Isajet{{\sc Isajet}}

\preprint{\vbox{OU-HEP-130801, KIAS-P13049}}

\title{Mixed axion/neutralino dark matter\\
in the SUSY DFSZ axion model
}
\author{Kyu Jung Bae$^{a}$, Howard Baer$^{a}$ and Eung Jin Chun$^{b}$\\
$^a$Dept.\ of Physics and Astronomy, University of Oklahoma, Norman, OK 73019, USA\\
$^b$Korea Institute for Advanced Study, Seoul 130-722, Korea\\
E-mail: \email{bae@nhn.ou.edu}, \email{baer@nhn.ou.edu}, \email{ejchun@kias.re.kr}
}

\abstract{
We examine mixed axion/neutralino cold dark matter production in the SUSY DFSZ
axion model where an axion superfield couples to Higgs superfields.
We calculate a wide array of axino and saxion decay modes along with their
decay temperatures, and thermal and non-thermal production rates.
For a SUSY benchmark model with a standard underabundance (SUA) of Higgsino-like dark matter (DM), we find for the PQ scale $f_a\alt 10^{12}$ GeV that the DM abundance is mainly comprised of 
axions as the saxion/axino decay occurs before the standard neutralino freeze-out and thus its
abundance remains suppressed. For $10^{12}\lesssim f_a\lesssim 10^{14}$ GeV, the
saxion/axino decays occur after neutralino freeze-out so that the neutralino abundance is enhanced
by the production via decay and subsequent re-annihilation.
For $f_a \gtrsim 10^{14}$ GeV, both neutralino dark matter and dark radiation are
typically overproduced. For judicious parameter choices, these can be suppressed
and the combined neutralino/axion abundance brought into accord with measured values.
A SUSY benchmark model with a standard overabundance (SOA) of bino DM is also examined and 
typically remains excluded due at least to too great a neutralino DM abundance for
$f_a\lesssim 10^{15}$ GeV. For $f_a\gtrsim 10^{15}$ GeV and lower saxion masses, large entropy
production from saxion decay can dilute all relics and the SOA model can be
allowed by all constraints.
}

\begin{document}

\section{Introduction}
\label{sec:intro}

The recent discovery of a Higgs boson at the LHC
seemingly completes the discovery program for all matter states predicted by
the Standard Model (SM). And yet the SM in the present form is beset by two problems --
the strong CP problem in the QCD sector and the instability of scalar fields
under quantum corrections (the infamous quadratic divergences) in the electroweak sector.
The first of these can be solved by introducing a Peccei-Quinn (PQ) symmetry~\cite{pq}
and a concomitant {\it axion} field $a$~\cite{ww}.
The PQ symmetry is broken at a scale $f_a$ typically taken to be in the
range $f_a/N_{\rm DW}\sim 10^9-10^{12}$ GeV~\cite{ksvz,dfsz}.
Here $N_{\rm DW}$ is the domain wall number, which is 6 for the DFSZ model.
By introducing the new PQ scale $f_a$ into the model, one might
then expect the new scalar mass $m_h$ to blow up to at least the PQ scale.
The Higgs mass can be stabilized by introducing softly broken supersymmetry (SUSY),
where the soft SUSY breaking (SSB) terms are expected to be of order
the gravitino mass $m_{3/2}$ in gravity-mediated SUSY breaking models~\cite{nilles}.
In this case, the axion is but one element of an axion chiral superfield $A$ which
necessarily also includes an $R$-parity-even spin-zero saxion $s$ and an
$R$-parity-odd spin-$1/2$ axino $\ta$.
In gravity-mediation, the saxion is expected to obtain a SSB mass $m_s\sim m_{3/2}$.
The axino is also expected to obtain a mass $m_{\ta}\sim m_{3/2}$
unless special circumstances arise~\cite{gy,ckn,cl}.\footnote{Such a heavy axino cannot
be a dark matter candidate as it is overproduced by thermal scattering as will be discussed later.}
For $R$-parity conserving SUSY models -- as motivated by the need for proton stability --
the dark matter is then expected to consist of both
an axion and the lightest SUSY particle (LSP), {\it i.e. two dark matter particles}.
The LSP in gravity-mediation, which is assumed here,
is typically the lightest neutralino $\tz_1$, a WIMP candidate.
Thus, in this class of models, it is conceivable that both a WIMP and an axion
might be detected in dark matter search experiments.

To assess dark matter detection prospects, one must calculate the ultimate
abundance of both axions and WIMPs.
The calculation is considerably more involved than in the axion-only~\cite{vacmis,vg1}
or the WIMP-only case~\cite{gjk}.
In the PQ augmented Minimal Supersymmetric Standard Model (PQMSSM),
one may produce WIMPs thermally, but also
non-thermally via production and subsequent decay of both axinos and saxions.
In addition, late decay of saxions and axinos into SM particles after WIMP freeze-out
but before onset of Big-Bang Nucleosynthesis (BBN) can inject entropy and thus dilute all relics
present at the time of decay.
Thus, the ultimate axion/WIMP abundance also depends on the production and decay
of both saxions and axinos in the early universe.

The axion-axino-saxion kinetic terms and self-couplings
(in four component notation) are of the form
\be
{\cal L}=\left(1+\frac{\sqrt{2}\xi}{v_{PQ}}s\right)
\left[\frac{1}{2}\partial^\mu a\partial_\mu a+\frac{1}{2}\partial^\mu s\partial_\mu s
+\frac{i}{2}\bar{\ta}\dsl \ta\right]
\ee
where $\xi =\sum_i q_i^3v_i^2/v_{PQ}^2$. Here $q_i$ and $v_i$ denote PQ charges and
vacuum expectation values of PQ fields $S_i$, and the PQ scale $v_{PQ} = f_a/\sqrt{2}$ 
is given by $v_{PQ}=\sqrt{\sum_i q_i^2v_i^2}$.
In the above interaction, $\xi$ is typically $\sim 1$, but in some cases can be as small
as $\sim 0$~\cite{cl}.

The axino/saxion production/decay rates are model-dependent.
In the SUSY KSVZ case, where heavy quark superfields $Q$ and $Q^c$ are introduced
to implement the PQ symmetry, the axion supermultiplet couples to QCD gauge fields via a high-dimensional interaction which leads to thermal production rates~\cite{axino}
depending on the reheat temperature $T_R$.
Mixed axion/neutralino dark matter production in the SUSY KSVZ model has been
computed in Ref's~\cite{ckls,blrs,bls} for
the case of suppressed saxion coupling to axions, and in Ref.~\cite{bbl} for
unsuppressed couplings which lead to production of dark radiation from $s\to aa$
decay.\footnote{Further references for dark radiation from SUSY axion models
include~\cite{gs2,gs3,conlon,hasenkamp}.}

In the SUSY DFSZ model, no exotic quark superfields are needed as 
the  Higgs doublet superfields, $H_u$ and $H_d$, are assumed to carry PQ charges.
An attractive feature of the DFSZ model is that it provides a simple
resolution of the so-called SUSY $\mu$ problem~\cite{kn}:
why is the superpotential $\mu$ term at the $m_{3/2}$ scale
(as required by phenomenology) instead of as high as the (reduced) Planck scale
$M_P\simeq 2\times 10^{18}$ GeV, as expected for SUSY preserving terms?
In the SUSY DFSZ model, the $\mu$ term is forbidden at tree-level
by the PQ symmetry. However, a superpotential term such as
\be
W\ni \lambda \frac{{S}^2}{M_P}{H}_u{H}_d
\ee
can be allowed.
After the PQ symmetry breaking by a vacuum expectation value of the scalar component of $S$, $\langle S\rangle\sim f_a$,
a mu term
\be \mu\sim \lambda f_a^2/M_P
\ee
will be induced.
The mu term is then at or around the weak scale
for $f_a\sim 10^{10}-10^{11}$ GeV assuming $\lambda\sim 1$.

The axion supermultiplet in DFSZ model couples directly to the Higgs fields
with an interaction given by
\be
{\cal L}_{{\rm DFSZ}}=\int d^2\theta (1+B\theta^2)\mu e^{c_H  A/v_{PQ}}  H_u  H_d,
\label{eq:superptl}
\ee
where $1+B \theta^2$ is a SUSY breaking spurion field and $c_H$ is the PQ charge of the 
Higgs bilinear operator $H_u H_d$.

The production and decay channels of saxions and axinos are very different in the SUSY DFSZ case
as compared to SUSY KSVZ.
Due to the renormalizable DFSZ interactions, thermal production rates
for axinos/saxions are independent of $T_R$.
In addition, for given $v_{PQ}$, saxion and axino decay rates are larger and
there are many more decay final states as compared to SUSY KSVZ.
As a result, for comparable values of masses and $f_a$,
the DFSZ saxion and axino are expected to be much shorter lived as
compared to the KSVZ case.

Dark matter production in the SUSY DFSZ model has been considered previously.
In Ref.~\cite{chun, bci11}, the overall WIMP production scenario for SUSY DFSZ was portrayed.
In Ref.~\cite{bci}, detailed calculations of axino production and decay were included.
In the present work, we augment these previous studies by including further
axino decay modes along with detailed computations of saxion decay rates.
We examine not only thermal production of axinos but also thermal and non-thermal
production of saxions. Finally, we account for production of axions as well, which are
necessarily present and add to the predicted dark matter abundance.

Our results also depend on which particular SUSY model spectrum is assumed.
We introduce in Sec.~\ref{sec:bm} two SUSY benchmark models
(the same points as in Ref.~\cite{bbl} for ease of comparison with the KSVZ case):
one with a bino-like LSP and a standard overabundance of WIMP dark matter (SOA)
and one with a Higgsino-like LSP (as motivated by recent naturalness studies~\cite{rns}),
which contains a standard underabundance of Higgsino-like WIMP dark matter (SUA).
A concise summary of our results for the SUA case has been presented earlier in Ref.~\cite{prl};
in the present work, we provide detailed discussion and formulae, and also consider the SOA
case.
In Sec.~\ref{sec:sdecay}, we present simplified formulae for the saxion decay widths
and exact leading order branching fractions and decay temperatures $T_D^s$.
In Sec.~\ref{sec:axdecay} we present similar results for axino decays.
In Sec.~\ref{sec:prod}, we briefly discuss axion production and
thermal axino and thermal/non-thermal saxion production rates.
We evaluate under which conditions axinos or saxions can
temporarily dominate the matter density of the universe.
In Sec.~\ref{sec:cosmo}, we examine several cosmological scenarios for the
SUA and SOA benchmarks:
1.\ low $(f_a\sim 10^{10}-10^{12}$ GeV),
2.\ medium ($f_a\sim 10^{12}-10^{14}$ GeV), and
3.\ high ($f_a\sim 10^{14}-10^{16}$ GeV)
ranges of the PQ scale.
While our SUA benchmark point easily lives in the low $f_a$ regime, it can with trouble
also be accommodated at medium and high $f_a$ values. In contrast, the SOA
benchmark fails to be viable at low or medium $f_a$, but can be viable at very high
$f_a\gtrsim 10^{15}$ GeV under certain restrictions such as a low enough $m_s$ value such that
saxion decays to sparticles are kinematically disallowed.
This latter point is especially important in that far higher $f_a$ values
can be accommodated in SUSY axion models
than are usually considered from non-SUSY models:
this is possible due to the capacity for large entropy
dilution along with the usual possibility of a small initial
axion misalignment angle~\cite{bl}.
In an Appendix, we list exact leading order saxion and axino decay formulae for the
DFSZ SUSY axion model.

\section{Two benchmark models for DFSZ SUSY study}
\label{sec:bm}

In this Section, we summarize two SUSY model benchmark points which are useful
for illustrating the dark matter production in the SUSY DFSZ axion model:
one (labeled as SUA) has a standard thermal
{\it underabundance} of neutralino cold dark matter (CDM) while the other (labeled as SOA)
has a standard thermal {\it overabundance} of neutralinos.

The first point-- listed as SUA-- comes from
{\it radiatively-driven natural SUSY}~\cite{rns}
with parameters from the 2-parameter non-universal Higgs model
\be
\mbox{$(m_0,\ m_{1/2},\ A_0,\ \tan\beta )$ = $(7025\ {\rm GeV},\ 568.3\ {\rm GeV},\ -11426.6\ {\rm GeV},\ 8.55)$}
\ee
with input parameters $(\mu,\ m_A)=(150,\ 1000)$ GeV.
We generate the SUSY model spectra with Isajet 7.83~\cite{isajet}.
As shown in Table~\ref{tab:bm},
with $m_{\tg}=1.56$ TeV and $m_{\tq}\simeq 7$ TeV, it is safe from LHC searches.
It has $m_h=125$ GeV and a Higgsino-like neutralino with mass $m_{\tz_1}=135.4$ GeV
and standard thermal abundance from IsaReD~\cite{isared} of
$\Omega_{\tz_1}^{MSSM}h^2=0.01$, low by a factor $\sim 10$ from the measured
dark matter density.
Some relevant parameters, masses and direct detection cross sections
are listed in Table~\ref{tab:bm}.
It has very low electroweak finetuning.

For the SOA case, we adopt the mSUGRA/CMSSM model with parameters
\be
 \mbox{$(m_0,\ m_{1/2},\ A_0,\ \tan\beta ,\ sign(\mu ))$ = $(3500\ {\rm GeV},\ 500\ {\rm GeV},\ -7000\ {\rm GeV},\ 10,\ +)$}
\ee
The SOA point  has $m_{\tg}=1.3$ TeV and $m_{\tq}\simeq 3.6$ TeV,
so it is just beyond current LHC sparticle search constraints.
It is also consistent with the LHC Higgs discovery since $m_h=125$ GeV.
The lightest neutralino is mainly bino-like with $m_{\tz_1}=224.1$ GeV, and the
standard neutralino thermal abundance  is found to be
$\Omega_{\tz_1}^{\rm MSSM}h^2=6.8$,
a factor of $\sim 60$ above the measured value~\cite{wmap9}.
Due to its heavy 3rd generation squark masses and large $\mu$
parameter, this point has very high electroweak finetuning~\cite{sug_ft}.

%
\begin{table}\centering
\begin{tabular}{lcc}
\hline
 & SUA (RNS2) & SOA (mSUGRA)   \\
\hline
$m_0$ & 7025 & 3500 \\
$m_{1/2}$  & 568.3& 500  \\
$A_0$ & -11426.6 & -7000  \\
$\tan\beta$  & 8.55 & 10  \\
$\mu $ & 150 & 2598.1 \\
$m_A$ & 1000 & 4284.2 \\
$m_h$ & 125.0 & 125.0 \\
$m_{\tg}$ & 1562 & 1312 \\
$m_{\tu}$ & 7021 & 3612 \\
$m_{\tst_1}$ & 1860 & 669 \\
$m_{\tz_1}$ & 135.4 & 224.1 \\
\hline
$\Omega^{\rm std}_{\tz_1} h^2$ & 0.01 & 6.8 \\
$\sigma^{\rm SI}(\tz_1 p)$ pb & $1.7\times 10^{-8}$ & $1.6\times 10^{-12}$ \\
\hline
\end{tabular}
\caption{Masses and parameters in~GeV units for two benchmark points
computed with \Isajet\,7.83 and using $m_t=173.2$ GeV.
}
\label{tab:bm}
\end{table}

\section{Decay of saxion}
\label{sec:sdecay}

In this section, we present simplified formulae for the partial decay widths of saxions.
These widths play an essential role in determining the cosmic densities of
mixed axion/neutralino cold dark matter.
Since the saxion mixes with the CP-even Higgs bosons $h$ and $H$,
it has similar decay channels via a tiny mixing coupling proportional to $\sim \mu/f_a$.
The couplings can be extracted by integrating Eq.~(\ref{eq:superptl}).
We list all the possible saxion decay channels in the following.

\begin{itemize}

\item $s\to hh~/~HH~/~hH~/~AA~/~H^+H^-$.

The saxion decays to pairs of Higgs states arise from the saxion
trilinear interaction as well as its mixing in Eq.~(\ref{eq:superptl}).
For a very heavy saxion, the mixing effect can be safely neglected
and the partial decay widths, neglecting phase space factors (these are included in the Appendix
and also in all the numerical results) are approximately given by
\begin{eqnarray}
\Gamma(s\to hh)&\approx&\frac{c_H^2}{16\pi}\frac{\mu^4}{v_{PQ}^2}
\left(1-\frac{m_A^2\cos^2\beta}{\mu^2}\right)^2\frac{1}{m_s},
\label{eq:approx_shh}\\
\Gamma(s\to HH)&\approx&\frac{c_H^2}{16\pi}\frac{\mu^4}{v_{PQ}^2}
\left(1+\frac{m_A^2\cos^2\beta}{\mu^2}\right)^2\frac{1}{m_s},
\label{eq:approx_sHH}\\
\Gamma(s\to hH)&\approx&
\frac{c_H^2}{32\pi}\left(\frac{m_A^2\cos\beta}{v_{PQ}}\right)^2\frac{1}{m_s},
\label{eq:approx_shH}\\
\Gamma(s\to H^+H^-)&\simeq& 2\times\Gamma(s\to AA)\approx
\frac{c_H^2}{8\pi}\frac{\mu^4}{v_{PQ}^2}
\left(1+\frac{m_A^2\cos^2\beta}{\mu^2}\right)^2\frac{1}{m_s}.
\label{eq:approx_sAA}
\end{eqnarray}
Note that we take the limit of decoupling-- $m_A^2\gg m_h^2$ and large $\tan\beta\gg1$-- unless otherwise stated.

\item $s\to ZZ~/~W^+W^-~/~ZA~/~W^+H^-$.

These decay modes arise from the mixing between the saxion and Higgs states.
For a heavy saxion, its decays into gauge boson states are dominated by
the decays into Goldstone states and thus
we can obtain similar approximate formulae as for the Higgs final states:
\begin{eqnarray}
\Gamma(s\to W^+W^-)&\simeq&2\times\Gamma(s\to ZZ)=
\frac{c_H^2}{8\pi}\frac{\mu^4}{v_{PQ}^2}
\left(1-\frac{m_A^2\cos^2\beta}{\mu^2}\right)^2\frac{1}{m_s},\\
\Gamma(s\to W^+H^-)&=&\Gamma(s\to W^-H^+)\simeq\Gamma(s\to ZA)\approx
\frac{c_H^2}{16\pi}\frac{m_A^4\cos^2\beta}{v_{PQ}^2}\frac{1}{m_s}.
\label{eq:approx_sVphi}
\end{eqnarray}

\item $s\to f\bar{f}$.

These modes are obvious due to the saxion mixing with Higgs states and their couplings contain 
a suppression factor of   $m_f/v_{PQ}$.
Thus, the decay rate is expected to be very small compared to the above decay modes
for generic parameter values with $m_f \ll \mu, m_A$.  For the case of $s \to t \bar t$ decay,
the decay rate is given by
\begin{equation}
\Gamma(s \to t \bar t) \approx {N_c \over 4 \pi} {c_H^2m_t^2 \over v_{PQ}^2}
\frac{\mu^4}{m_s^3}\left(1-\frac{m_A^2\cos^2\beta}{2\mu^2}\right)^2.
\end{equation}

\item $s\to \widetilde{Z}_i\widetilde{Z}_j~/~\widetilde{W}_i\widetilde{W}_j$.

In the heavy saxion limit, $m_s \gg \mu$, the saxion decays dominantly to Higgsino-like neutralinos and charginos whose
partial decay widths are given by
\begin{eqnarray}
\Gamma(s\to\mbox{all neutralinos})&\approx&\frac{c_H^2}{64\pi}\left(\frac{\mu}{v_{PQ}}\right)^2m_s,
\label{eq:approx_sneutneut}\\
\Gamma(s\to\mbox{all charginos})&\approx&\frac{c_H^2}{64\pi}\left(\frac{\mu}{v_{PQ}}\right)^2m_s.
\label{eq:approx_scharchar}
\end{eqnarray}

\item $s\to \tilde{f}\tilde{f}$.

Similarly to fermion modes, these decay rates are also expected to be very small due to the Yukawa suppression.
For the case of $s\to \tilde{t}_1\tilde{t}_1$,
\begin{equation}
\Gamma(s\to\tilde{t}_1\tilde{t}_1)\approx
\frac{N_c}{2\pi}\frac{c_H^2\mu^4}{v_{PQ}^2}\frac{m_t^4}{m_s^5}
\left(1-\frac{m_A^2\cos^2\beta}{2\mu^2}\right)^2.
\end{equation}
Note that we neglect the squark mixing effect which is not very large for our benchmark points.

\item $s\to aa~/~\ta\ta$.

Finally, the saxion has generic trilinear couplings to axions and axinos  which depend on the details of PQ symmetry breaking sector.
The partial decay widths are given by
\begin{eqnarray}
\Gamma(s\to aa)&=&\frac{\xi^2m_s^3}{64\pi v_{PQ}^2},\\
\Gamma(s\to\ta\ta)&=&\frac{\xi^2}{8\pi}\frac{m_{\ta}^2m_s}{v_{PQ}^2}\left(1-4\frac{m_{\ta}^2}{m_s^2}\right)^{3/2}.
\end{eqnarray}
Here the model dependent parameter $\xi \lesssim 1$ quantifies the axion superfield trilinear coupling.

\end{itemize}

In the following, we will show explicit numerical examples of saxion decays into the
aforementioned final states.
We can see the relative ratios of such decay modes for the SUA and SOA benchmark points
and for $\xi =0$ or 1.

\subsection{Saxion branching fractions}

\begin{figure}
\begin{center}
\includegraphics[height=6.9cm]{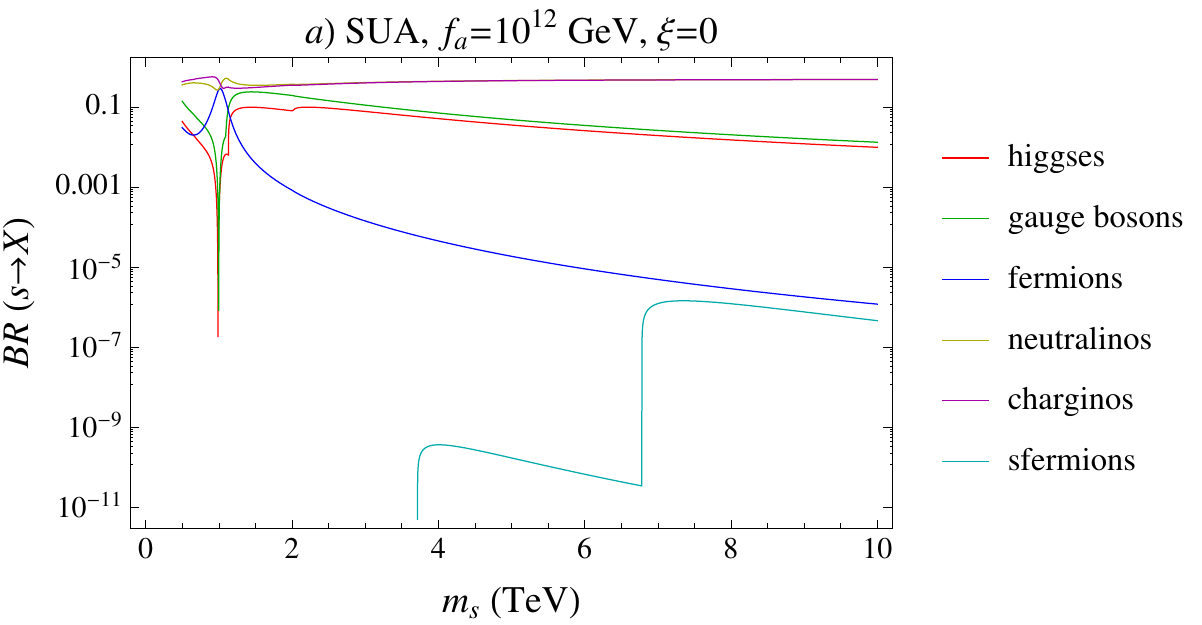}
\includegraphics[height=6.9cm]{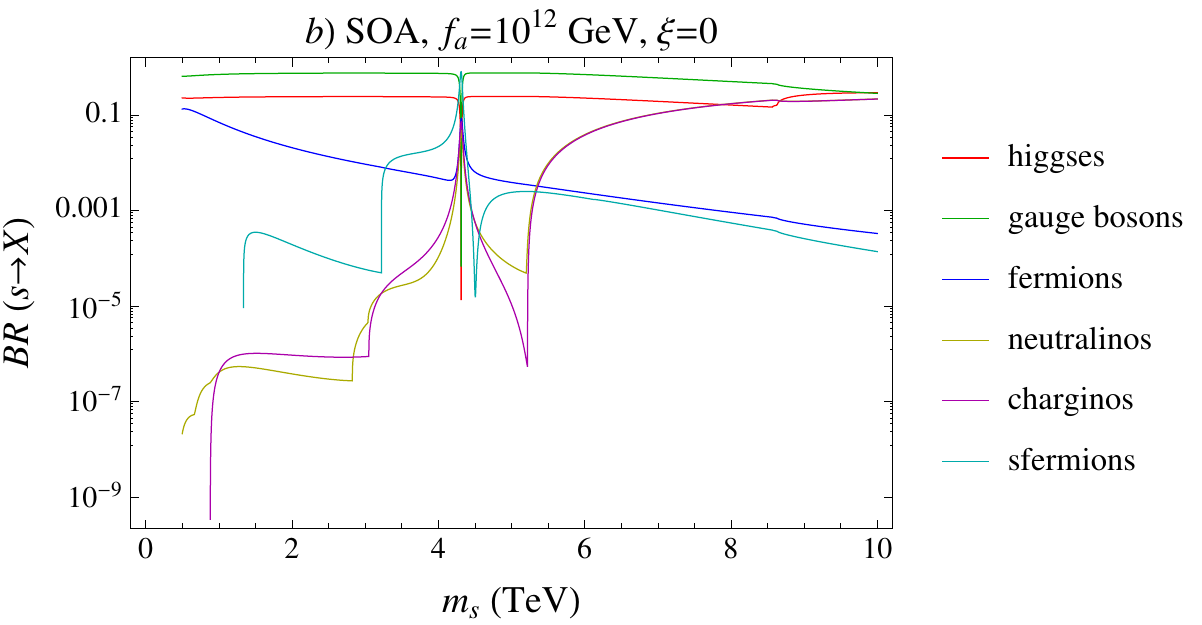}
\caption{Saxion decay branching ratios for $f_a=10^{12}$ GeV and $\xi=0$
for {\it a}) SUA and {\it b}) SOA benchmark points.
\label{fig:sax_BR_xi=0}}
\end{center}
\end{figure}

Fig.~\ref{fig:sax_BR_xi=0} shows saxion branching ratios (BR) versus $m_s$ for
the case of $\xi=0$ (for which there are no decays into axion or axino pairs)
for {\it a}) the SUA case and {\it b}) the SOA case.
We take $f_a=10^{12}$ GeV.
For $\xi =0$  and a large saxion mass in the SUA case, the most important decays are into SUSY particles:
charginos and neutralinos (the curves nearly overlap).
Decays into gauge and Higgs particles are subdominant-- about one or two orders of magnitude
smaller than the neutralino and chargino modes for multi-TeV values of $m_s$.
This behavior can be understood from the approximate formulae,
Eqs.~(\ref{eq:approx_shh})-(\ref{eq:approx_scharchar}).
The partial decay widths are proportional to $m_s$ for the decay to neutralinos and charginos
while they are inversely proportional to $m_s$ for the decays into gauge and Higgs states.
For smaller saxion mass, {\it e.g.} $m_s\lesssim 1.5$ TeV, the decay into top quark pairs also becomes sizable.
Note that the decays into gauge and Higgs states are strongly suppressed for the saxion mass around 1 TeV
for which the saxion-Higgs mixing is maximized so that the saxion coupling to gauge and Higgs particles become very small due to cancellation.

In frame {\it b}) we show the case for the SOA benchmark.
Here,  the dominant decay modes are instead into gauge boson and Higgs final states.
This behavior arises because for the SOA case $\mu\simeq 2.6$ TeV which is quite large.
From Eq's~(\ref{eq:approx_shh})-(\ref{eq:approx_sVphi}),
we can see that these partial widths are proportional to $\mu^4$ instead of $\mu^2$
as per the decay to -inos.

\begin{figure}
\begin{center}
\includegraphics[height=6.9cm]{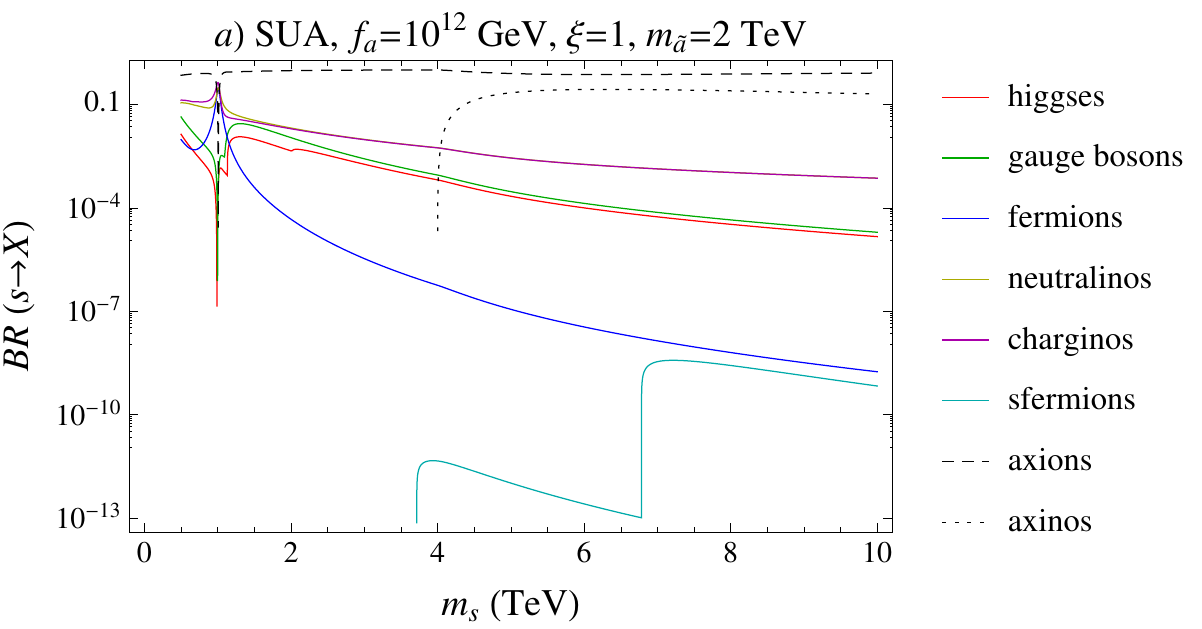}
\includegraphics[height=6.9cm]{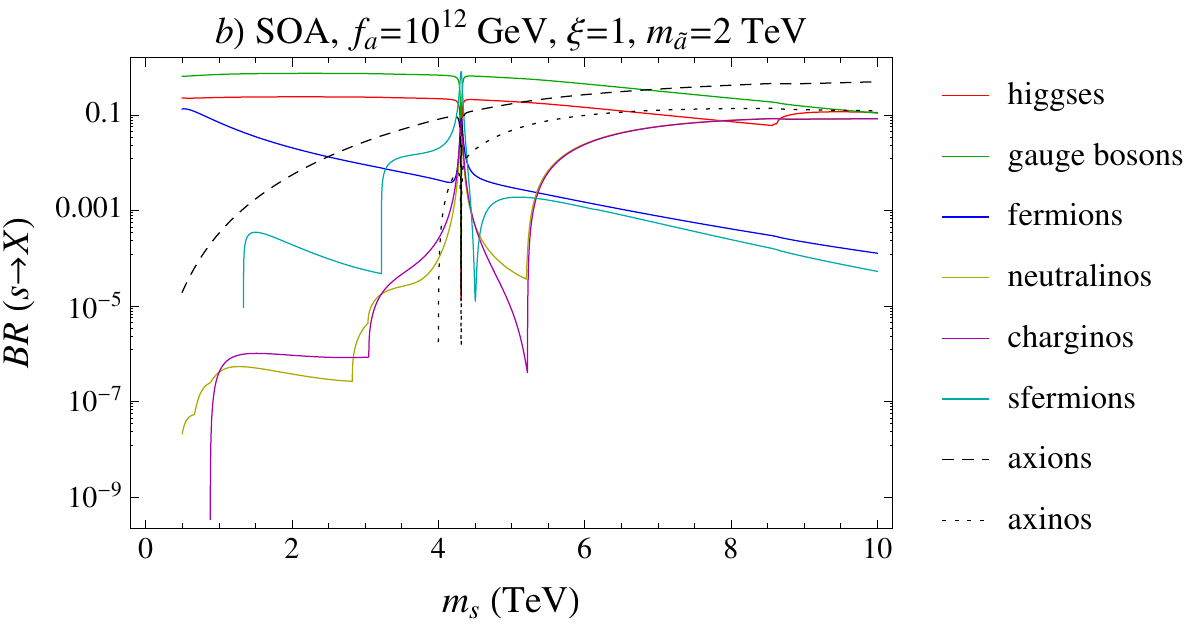}
\caption{Saxion branching ratios for $f_a=10^{12}$ GeV and $\xi=1$ for {\it a}) SUA and
{\it b}) SOA benchmark points.
Here we take $m_{\ta}=2$ TeV to show saxion decay into axino pair.
\label{fig:sax_BR_xi=1}}
\end{center}
\end{figure}

In Fig.~\ref{fig:sax_BR_xi=1}, we show the saxion BRs for $\xi=1$,
assuming $m_{\ta}=2$ TeV.
In the case of SUA, the most important mode is $s\to aa$
where the BR is a few orders of magnitude larger than other MSSM modes
including those to sparticles and SM particles for a large saxion mass.
This can be understood
since the decay into axion pairs is proportional to $m_s^3$ while the others are
proportional to $m_s$ or $1/m_s$.
When the saxion mass is much larger than the SUSY particle masses--
{\it i.e.} for saxion mass around 10 TeV--
$BR(s\to\mbox{SM})$ is smaller than $10^{-3}$
and thus the constraint from dark radiation becomes stronger
if saxions dominate the energy density of the universe.
Also, if $s\to\ta\ta$ is allowed, then it might become the dominant source of neutralino dark matter
via the axino decay.

For smaller saxion mass, $m_s\lesssim 1$ TeV,
the BRs of the MSSM channels become larger so that the constraint from dark radiation becomes relieved.
The saxion mass around 1 TeV shows an interesting behavior in that
 the neutralino/chargino modes become sizable and thus the amount of dark radiation 
from saxion decay can be drastically reduced.

In frame {\it b}), we show the $\xi=1$ case for the SOA benchmark.
For large $m_s\gtrsim 7$ TeV, the $s\to aa$  mode is again dominant.
It is worth noting that the decay widths to neutralinos and charginos never becomes larger than the
Higgs and gauge boson modes for $m_s\lesssim 10$ TeV in contrast to the SUA case.
Thus, augmentation of neutralino dark matter via late decay of saxions can not be very large,
which will be discussed in detail in Sec.~\ref{sec:prod}.
For $m_s\lesssim 7$ TeV, the dominant final state is into SM particles;
thus, in this case, low rates of dark radiation occur even when saxions dominate the universe.
We will discuss this in Section~\ref{sec:cosmo}.
For $m_s\simeq m_A$, mixing between saxions and Higgses becomes very large
so that the sfermion final states become the dominant decay modes.

\subsection{Saxion decay temperature}

In this subsection, we show saxion decay temperature values expected from the
SUA and SOA benchmarks for $\xi =0$ and $1$.
The temperature at which saxions decay is related to the total decay width via
\be
T_D =\sqrt{\Gamma M_P}/(\pi^2g_*(T_D)/90)^{1/4}
\ee
where $M_P$ is the reduced Planck mass and $g_*$ is the effective number of degrees of freedom
at temperature $T_D$.
Note that we assume a radiation-dominated universe in the temperature plots.
The case of a saxion-dominated universe will be discussed in Sec.~\ref{sec:prod}.

\begin{figure}
\begin{center}
\includegraphics[height=6.9cm]{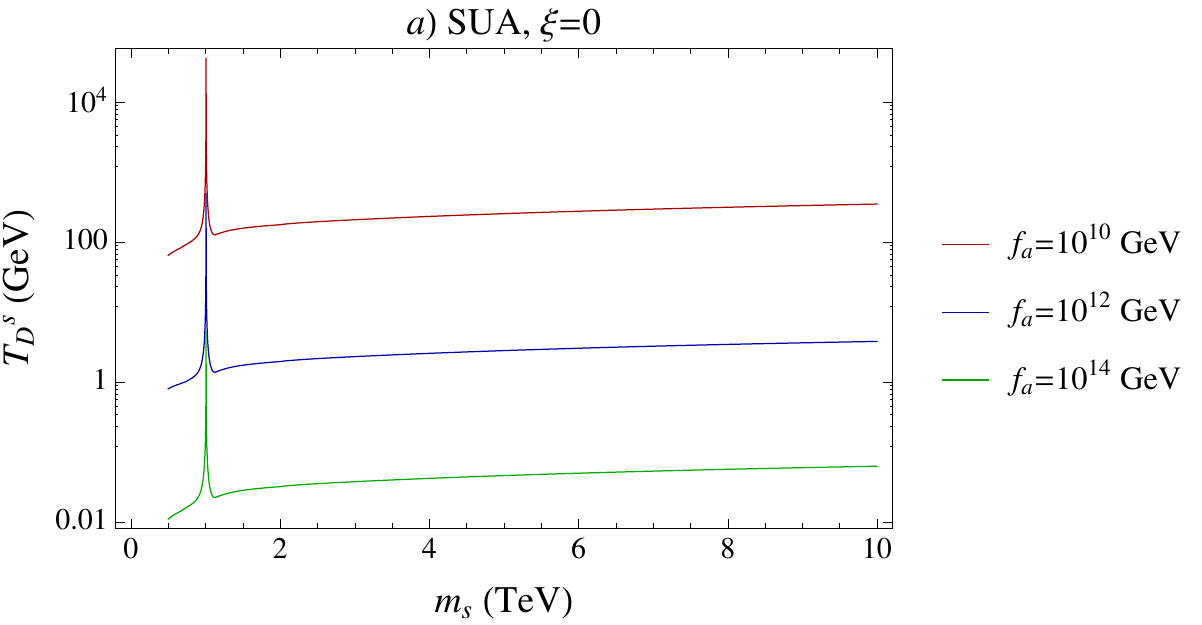}
\includegraphics[height=6.9cm]{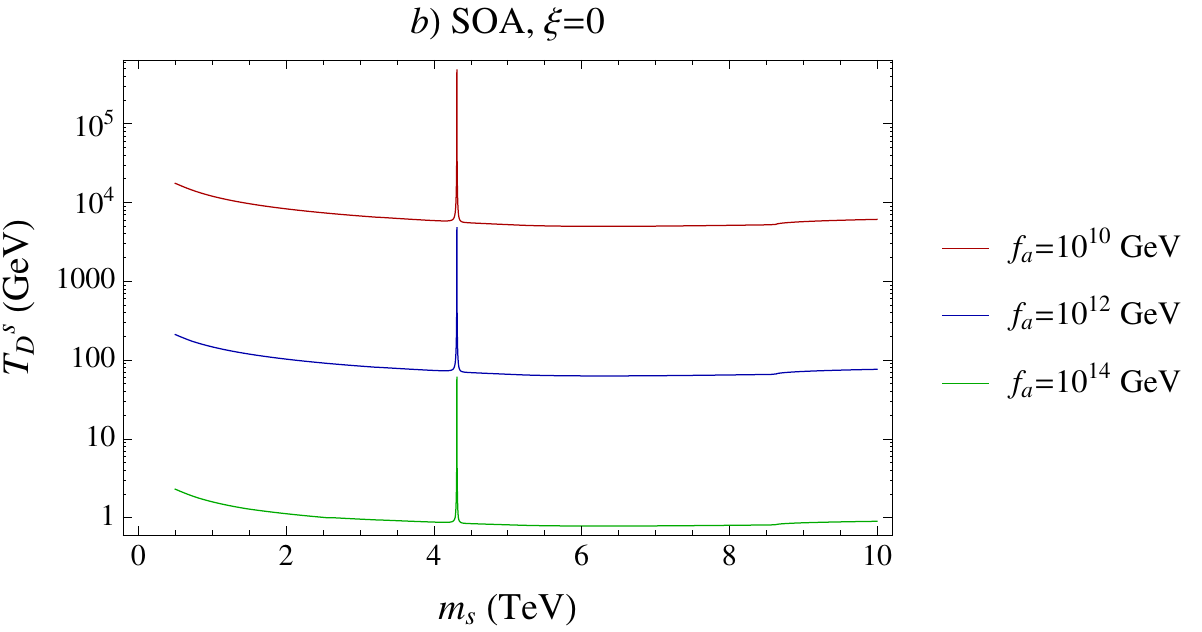}
\caption{Saxion decay temperature for $f_a=10^{12}$ GeV and $\xi=0$
for {\it a}) SUA and {\it b}) SOA benchmark points.
\label{fig:sax_TD_xi=0}}
\end{center}
\end{figure}

In Fig.~\ref{fig:sax_TD_xi=0}, we show the saxion decay temperature for
$f_a=10^{10}$, $10^{12}$ and $10^{14}$ GeV.
For the $\xi=0$ case shown in frame {\it a}),
$T_D^s\sim 10$ MeV for $f_a\sim 10^{14}$ GeV so that even for these large values
of $f_a$, the saxion decays before the onset of BBN.
For $f_a\sim 10^{10}$ GeV,
the decay temperature typically ranges up to 100 GeV. This is typically well above
WIMP freezeout temperature, given approximately by $T_{fr}\sim m_{\tz_1}/25$.
Thus, for low $f_a$, saxions in the DFSZ model tend to decay before freezeout,
and so the standard thermal WIMP abundance calculation may remain valid.

In the SOA case shown in frame {\it b}), $T_D^s$ varies from $1-10^4$ GeV as
$f_a$ ranges from $10^{14}-10^{10}$ GeV. Thus,
saxions tend to decay well before WIMP freezeout unless $f_a$ is as large as $10^{14}$ GeV,
in which case saxion decays are suppressed.

\begin{figure}
\begin{center}
\includegraphics[width=14cm]{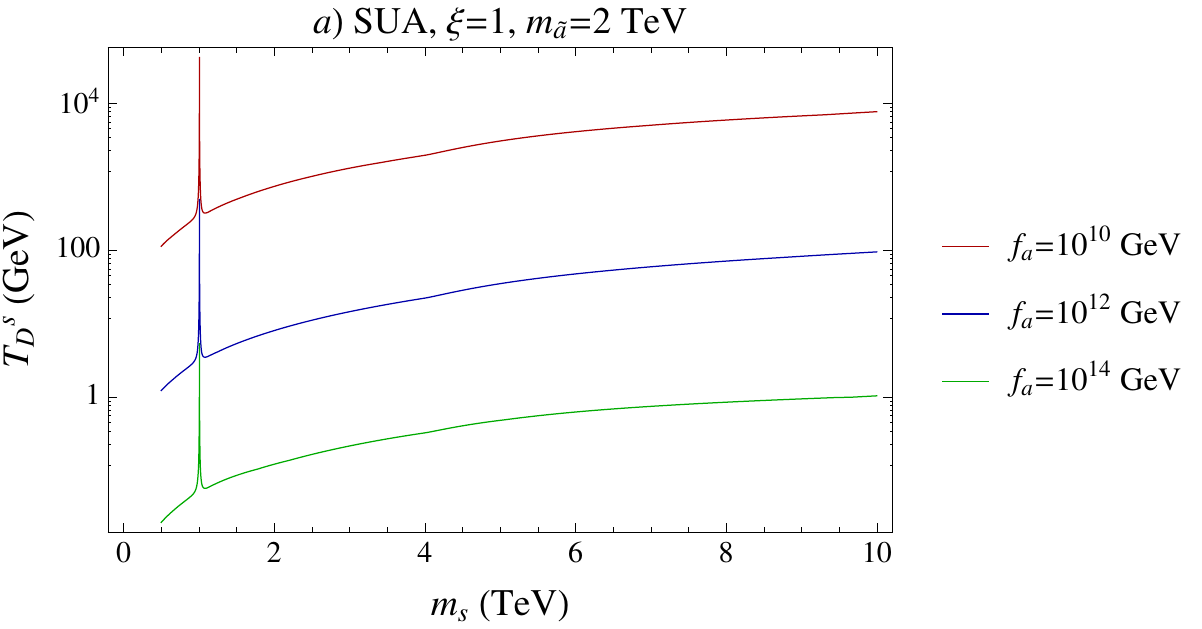}
\includegraphics[width=14cm]{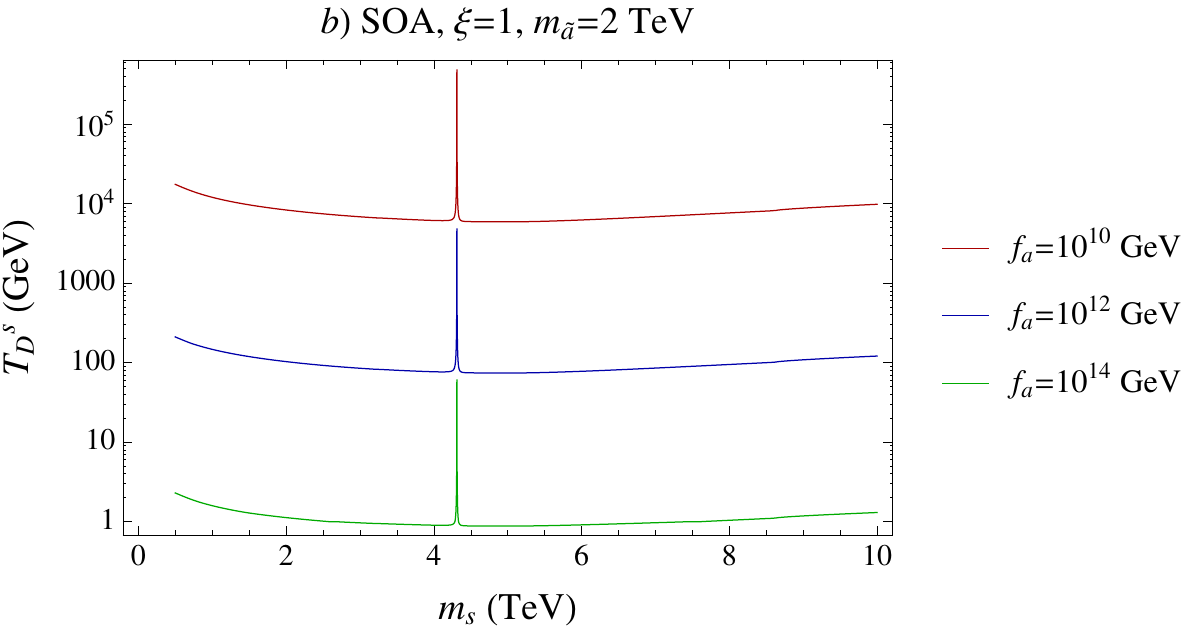}
\caption{Decay temperature for saxion with $f_a=10^{12}$ GeV,
$\xi=1$ and $m_{\ta}=2$ TeV for {\it a}) SUA and {\it b}) SOA
benchmark points.
\label{fig:sax_TD_xi=1}}
\end{center}
\end{figure}

In Fig.~\ref{fig:sax_TD_xi=1}, we show the saxion decay temperature for $\xi =1$.
In frame {\it a}) for the SUA case, and with $m_{\ta}=2$ TeV, the presence of the
$s\to aa,\ta\ta$ modes increases even further the saxion decay temperature compared
to the $\xi =0$ case. For $m_s=10$ TeV, we see that $T_D^s$ ranges from $1-10^4$ GeV
as $f_a\sim 10^{14}-10^{10}$ GeV. In frame {\it b}), we find a similar decay temperature
for SOA as compared to SUA since $T_D^s$ is dominated by the $s\to aa,\ \ta\ta$ widths
which are the same for both cases.

The upshot of this section is that in the DFSZ model, direct coupling of saxions to
Higgs and Higgsinos increases the saxion decay widths compared to the KSVZ model,
typically causing saxions to decay before the BBN onset even for $f_a$ as large as $10^{14}$
GeV, and for smaller $f_a$ values, saxions tend to decay even before neutralino freezeout,
thus leading to no augmentation of neutralino relic density via late-time reannihilation.

\section{Axino decays}
\label{sec:axdecay}

Similar to the saxion case, the axino trilinear couplings arise directly or indirectly 
through the axino-Higgsino mixing from the superpotental (\ref{eq:superptl}).
The possible decay modes include the following.
\begin{itemize}

\item $\ta \to \widetilde{Z}_i h~/~\widetilde{Z}_iH~/~\widetilde{Z}_iA$.

The decays into neutralinos and Higgs bosons come from the axino-Higgsino-Higgs interaction, 
so the dominant decay modes are into Higgsino-like neutralino states in the limit of heavy axino.
In the heavy axino limit ({\it i.e.} $m_{\ta}\gg\mu$), the partial decay width is given by
\begin{equation}
\Gamma (\ta\to\widetilde{Z}_i\phi ) \approx 2 \times 3\times \frac{c_H^2}{64\pi}\left(\frac{\mu}{v_{PQ}}\right)^2m_{\tilde{a}}.
\end{equation}
where $\phi=h,\ H$ and $A$.

\item $\ta\to\widetilde{W}_i^{\pm}H^{\mp}$.

These decay modes arise similarly to the previous ones.
For the heavy axino limit, the partial width is determined by
\begin{equation}
\Gamma (\ta\to\widetilde{W}_i^{\pm}H^{\mp}) \approx
\frac{c_H^2}{16\pi}\left(\frac{\mu}{v_{PQ}}\right)^2m_{\tilde{a}}.
\end{equation}

\item $\ta\to\widetilde{Z}_iZ~/~\widetilde{W}_i^{\pm}W^{\mp}$.

The axino decays to gauge bosons arise from the axino-neutralino mixing.
In the limit of heavy axino, the corresponding decay rates can be obtained by considering
the decays into the Goldstone modes as follows:
\begin{eqnarray}
\Gamma (\ta\to\widetilde{Z}_iZ~)&\approx&\frac{c_H^2}{32\pi}\left(\frac{\mu}{v_{PQ}}\right)^2m_{\tilde{a}},\\
\Gamma (\ta\to\widetilde{W}_i^{\pm}W^{\mp})&\approx&\frac{c_H^2}{16\pi}\left(\frac{\mu}{v_{PQ}}\right)^2m_{\tilde{a}}.
\end{eqnarray}

\item $\ta\to\tilde{f}f$.

These modes also arise from axino-neutralino mixing.
In most cases, they are suppressed by $m_f/v_{PQ}$ as in the saxion case, 
and kinematically disallowed in most of parameter space with heavy matter scalars.
For the case of $\ta\to \tilde{t}_1\bar{t}+{\rm c.c.}$,
\begin{equation}
\Gamma(\ta\to\tilde{t}_1\bar{t}+\bar{\tilde{t}}_1 t)\approx
\frac{N_c}{32\pi}\frac{c_H^2\mu^4}{v_{PQ}^2}\frac{m_t^2}{m_{\ta}^3}
\left(1+\frac{m_{\ta}}{\mu\tan\beta}\right)^2.
\end{equation}
Note that we neglect the mixing effect of stop as in the case of $s\to\tilde{t}_1\tilde{t}_1$.

\end{itemize}

\subsection{Axino branching fractions}

\begin{figure}
\begin{center}
\includegraphics[height=6.9cm]{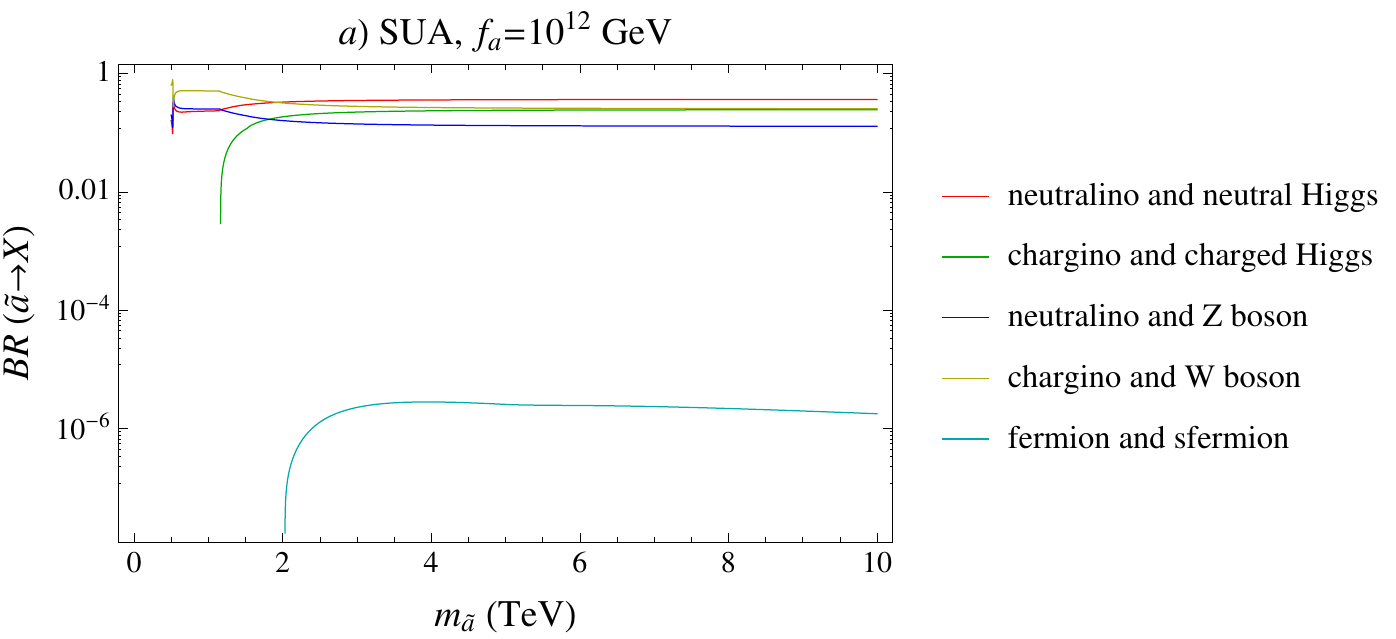}
\includegraphics[height=6.9cm]{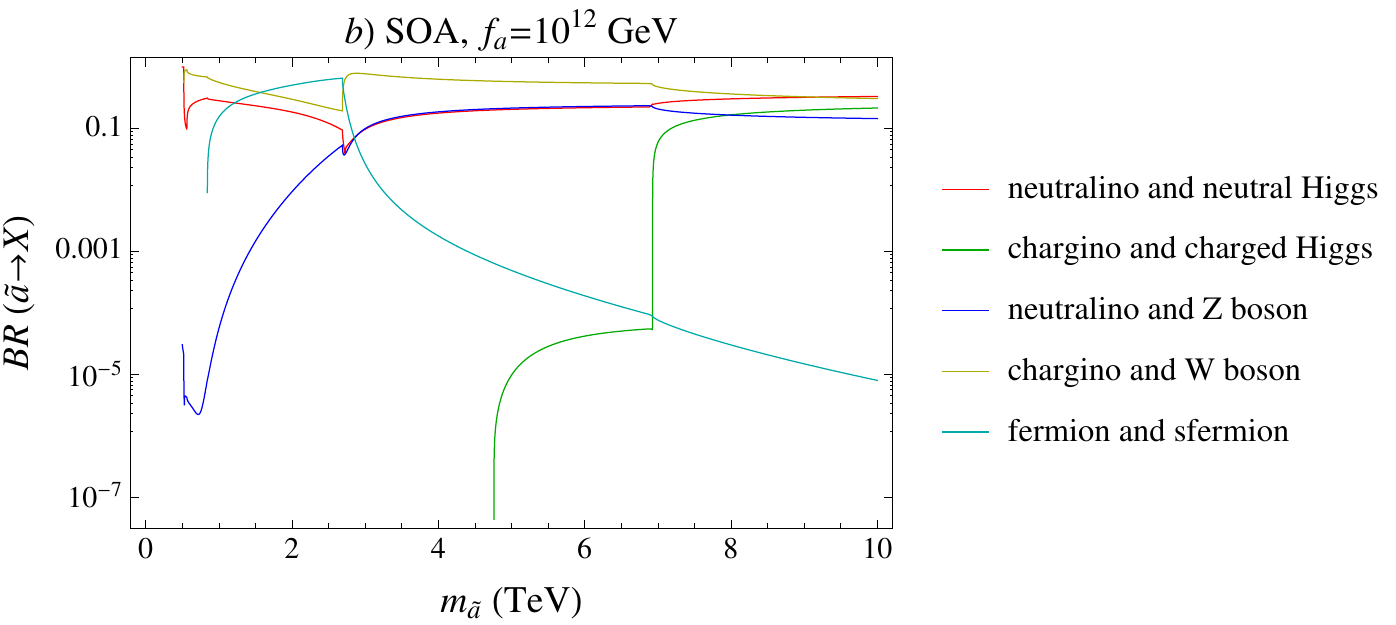}
\caption{Axino branching ratios for $f_a=10^{12}$ GeV and for {\it a}) SUA and
{\it b}) SOA benchmark points.
\label{fig:axno_BR}}
\end{center}
\end{figure}

In the Fig.~\ref{fig:axno_BR}, we show the axino branching fractions as a function of
$m_{\ta}$ for {\it a}) the SUA and {\it b}) the SOA benchmark points.
In most of parameter space, the branching fractions for decay to
neutralino+neutral Higgs, chargino+charged Higgs, neutralino+$Z$ and chargino$+W$ are
all comparable, in the tens of percent, while decays to fermion+sfermion are suppressed.
This is consistent with the above discussion and approximate formulae except for the region of
$m_{\ta}\sim m_{\widetilde{Z}_i}$ for which axino-neutralino mixing is enhanced.
For the SOA case, the qualitative features for large $m_{\ta}$ are almost the same as SUA case.
The differences come only from the different particle mass spectrum.
For $m_A\simeq\mu=2.6$ TeV, we can see the effect of maximized axino-Higgsino mixing
leading to the domination of the sfermion plus fermion mode.

\subsection{Axino decay temperature}

\begin{figure}
\begin{center}
\includegraphics[height=6.9cm]{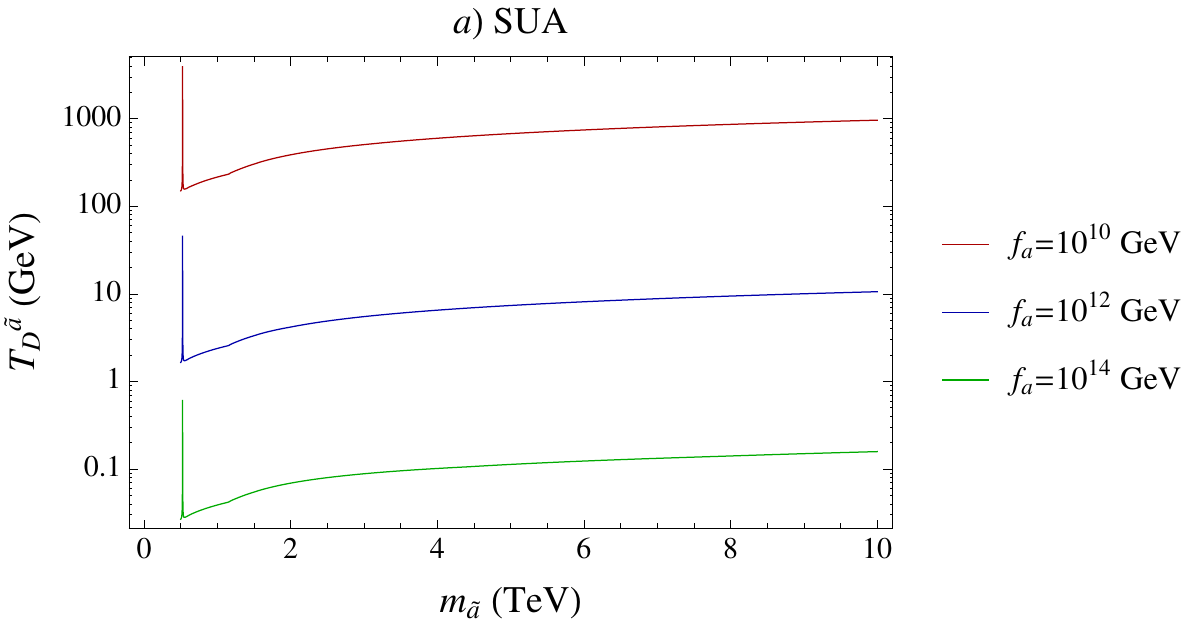}
\includegraphics[height=6.9cm]{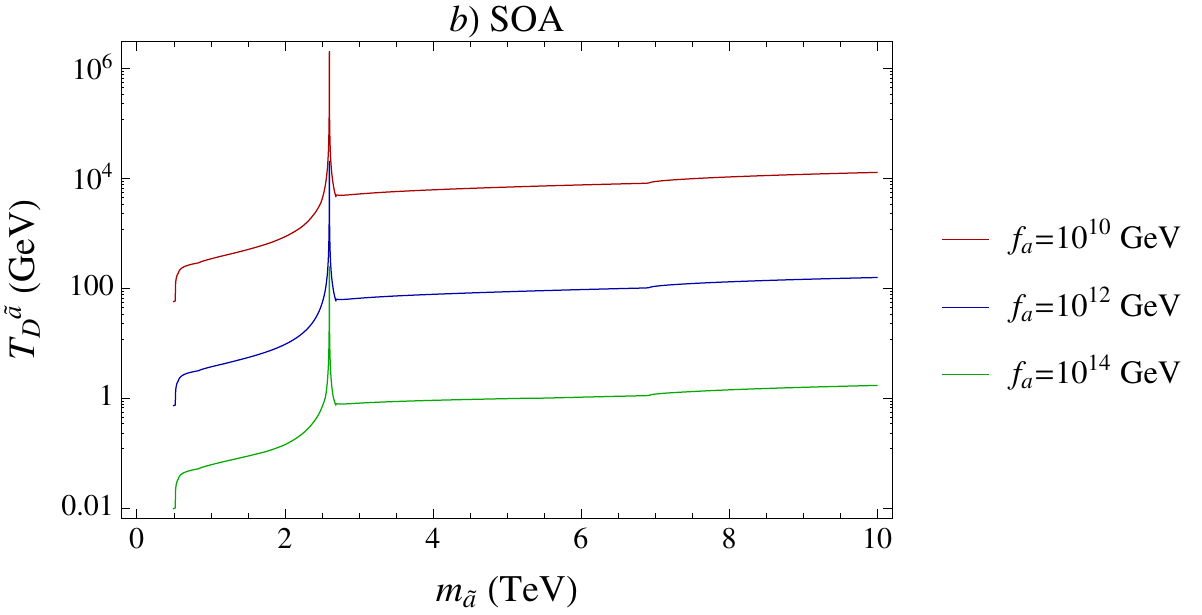}
\caption{Axino decay temperature for $f_a=10^{12}$ GeV and for {\it a}) SUA and
{\it b}) SOA benchmark points.
\label{fig:axno_TD}}
\end{center}
\end{figure}

In Fig.~\ref{fig:axno_TD}, we show the axino decay temperature $T_D^{\ta}$ versus
$m_{\ta}$ for {\it a}) the SUA and {\it b}) the SOA benchmarks, for
$f_a=10^{10}$, $10^{12}$ and $10^{14}$ GeV.
In SUA case, we see that $T_D^{\ta}$ varies from ${\cal O}(10)$ MeV to ${\cal O}(1)$ TeV depending
on the $f_a$ and $m_{\ta}$ values.
Since the axino always decays to SUSY particles, it should always augment the neutralino
abundance unless $T_D^{\ta}>T_{fr}$, in which case the usual thermal abundance applies, unless
affected by saxion decays. One typically has $T_D^{\ta}>T_{fr}$ as long as $f_a\lesssim 10^{11}$ GeV.
For the SOA case with a large $\mu =2.6$ GeV,
$T_D^{\ta}$ is about an order of magnitude higher than for the SUA case.

\section{Axion, axino and saxion production in the SUSY DFSZ model}

\label{sec:prod}

In this section, we will present formulae for the axion, axino and saxion production, and
discuss the possibility of saxion or axino domination in the early universe.

\subsection{Axion production}

Here we will assume the scenario where the PQ symmetry breaks before the end of inflation,
so that a nearly uniform value of the axion field $\theta_i\equiv a(x)/f_a$ is expected throughout the universe.
From the axion equation of motion, the axion field stays relatively
constant until temperatures approach the QCD scale $T_{\rm QCD}\sim 1$ GeV.
At this point, a temperature-dependent axion mass term turns on, and a
potential is induced for the axion field.
At temperature $T_a$ the axion field begins to oscillate,
filling the universe with low energy (cold) axions.
The standard axion relic density (via this vacuum misalignment mechanism)
is derived assuming that coherent oscillations begin in a radiation-dominated
universe and is given by~\cite{vacmis,vg1}
\be
\Omega_a^{\rm std} h^2\simeq 0.23 f(\theta_i)\theta_i^2
\left(\frac{f_a/N}{10^{12}\ {\rm GeV}}\right)^{7/6}
\label{eq:Oh2axionstd}
\ee
where $0< \theta_i<\pi$ and $f(\theta_i)$ is the anharmonicity
factor. Visinelli and Gondolo~\cite{vg1} parameterize the latter as
$f(\theta_i)=\left[\ln\left(\frac{e}{1-\theta_i^2/\pi^2}\right)\right]^{7/6}$.
The uncertainty in $\Omega_a h^2$ from vacuum misalignment is estimated
as plus-or-minus a factor of three.
If the axion field starts to oscillate during the matter dominated (MD) or the
decaying particle dominated (DD) phase ($T_D < T_a < T_e$), the axion relic
density will no longer be given by Eq.~(\ref{eq:Oh2axionstd}). The appropriate
expressions for each of these cases are given in the Appendix of Ref.~\cite{blrs}.

\subsection{Axino and saxion production}

In the case of the KSVZ models, thermal production of saxions and axinos is due to the anomaly interaction
of dimension 5:
\begin{eqnarray}
{\cal L}_{{\rm anomaly}}&=&-\frac{\sqrt{2}\alpha_s}{8\pi f_a/N_{\rm DW}}\int d^2\theta AW^aW^a+\mbox{h.c.}\nonumber\\
&=&\frac{\alpha_s}{8\pi f_a/N_{\rm DW}}\left(sG^a_{\mu\nu}G^{a\mu\nu}-i\bar{\ta}\frac{[\gamma^{\mu},\gamma^{\nu}]}{2}\gamma_5\tilde{g}^aG_{\mu\nu}^a+\cdots\right).
\end{eqnarray}
This higher dimensional couplings lead to thermally produced saxion and axino densities which are proportional
to the reheat temperature $T_R$.

In contrast, the axion supermultiplet in the SUSY DFSZ model has Yukawa-type (dimension 4)
interactions as shown in Eq.~(\ref{eq:superptl}).
As a consequence, the most important contributions for the saxion and axino production
arise near the kinematic thresholds of scattering processes leading to thermal production densities which are independent of
$T_R$ so long as $T_R$ is larger than the kinematic threshold for the specific process.
As was studied in Ref's~\cite{chun,bci11,bci}, the saxion and axino abundances
from thermal production are given by
\begin{eqnarray}
Y_{s}^{{\rm TP}}&=&10^{-7}\zeta_s\left(\frac{B\mu/\mu\mbox{ or }\mu}{\mbox{TeV}}\right)^2\left(\frac{10^{12}\mbox{ GeV}}{f_a}\right)^2,\\
Y_{\ta}^{{\rm TP}}&=&10^{-7}\zeta_{\ta}\left(\frac{\mu}{\mbox{TeV}}\right)^2\left(\frac{10^{12}\mbox{ GeV}}{f_a}\right)^2,
\end{eqnarray}
where $\zeta_s$ and $\zeta_{\ta}$ are model-dependent constants of order unity.
\footnote{For numerical discussion in Sec. \ref{sec:cosmo}, we take $\xi_s=\xi_{\ta}=1$ and max$[B\mu/\mu,\mu]$.}
Barring a specific model-dependence on  $\zeta_s$ and $\zeta_{\ta}$, we will
now examine the possibility of having a cosmological era dominated by thermally produced saxions or axinos.
For this, we need to compare two important quantities; the decay temperature $T_D$ and the saxion/axino-radiation equality temperature $T_e$.

Let us first discuss the saxion case.
The saxion-radiation equality temperature is given by
\begin{equation}
T_e^s=\frac43m_sY^{{\rm TP}}_s=\frac43 \times 10^{-7}\zeta_s\left(\frac{B\mu/\mu\mbox{ or }\mu}{\mbox{TeV}}\right)^2\left(\frac{10^{12}\mbox{ GeV}}{f_a}\right)^2m_s,
\end{equation}
and the decay temperature is
\begin{equation}
T_D^s=\sqrt{\Gamma_sM_p}\left(\frac{90}{\pi^2g_*}\right)^{1/4}.
\end{equation}
The saxion decay width is approximately given by
\begin{equation}
\Gamma_s\approx \frac{c_H^2}{32\pi}\left(\frac{\mu}{v_{PQ}}\right)^2m_s=\frac{c_H^2}{16\pi}\left(\frac{\mu}{f_a}\right)^2m_s.
\end{equation}
Here we only consider the decay width for $\xi=0$ and the dominant decays into neutralinos and charginos for the SUA scenario.
For the case of SOA, the situation does not significantly change.
The condition for the saxion domination, $T_e^s>T_D^s$, leads to
\begin{equation}
\frac{(B\mu/\mu\mbox{ or }\mu)^2/\mu}{f_a}\gtrsim 3\times10^{-5}
\left(\frac{c_H}{\zeta_s}\right)
\left(\frac{90}{g_*}\right)^{1/4}
\left(\frac{\rm TeV}{m_s}\right)^{1/2}.
\label{eq:TP_sax_dom}
\end{equation}
This condition is hardly achieved for $f_a\gtrsim10^9$ GeV unless $B\mu/\mu$ or $\mu$ is as large as 100 TeV.
Thus, we conclude that the saxion domination is unlikely to occur in the case of thermal production.
If $\xi\neq 0$, the saxion decay temperature becomes larger, and thus
the saxion domination is even less probable.  The same conclusion can be drawn for the axino case where
the axino domination requires
\begin{equation}
\frac{\mu}{f_a}\gtrsim 8\times10^{-5}
\left(\frac{c_H}{\zeta_{\ta}}\right)
\left(\frac{90}{g_*}\right)^{1/4}
\left(\frac{\rm TeV}{m_{\ta}}\right)^{1/2}
\end{equation}
which is also hard to meet.

\medskip

Next we consider saxion coherent oscillations. 
In this case, the saxion abundance is given by
\begin{equation}
Y_s^{\rm CO}=1.9\times10^{-6}\left(\frac{\rm GeV}{m_s}\right)
\left(\frac{{\rm min}[T_R,T_s]}{10^7\mbox{ GeV}}\right)
\left(\frac{f_a}{10^{12}\mbox{ GeV}}\right)^2
\end{equation}
assuming an initial saxion field amplitude of $s_0=f_a$.
From this, one finds that  the saxion domination occurs for
\begin{equation}
f_a\gtrsim 8\times 10^{13}\mbox{ GeV}\times c_H^{1/3}
\left(\frac{90}{g_*}\right)^{1/12}
\left(\frac{10^7\mbox{ GeV}}{{\rm min}[T_R,T_s]}\right)^{1/3}
\left(\frac{\mu}{150\mbox{ GeV}}\right)^{1/3}
\left(\frac{m_s}{\rm 5 TeV}\right)^{1/6}.
\label{eq:sax_dom_cond}
\end{equation}
Note that the lower limit on $f_a$ becomes larger for smaller $T_R$ or $T_s$.

\begin{figure}
\begin{center}
\includegraphics[height=6.9cm]{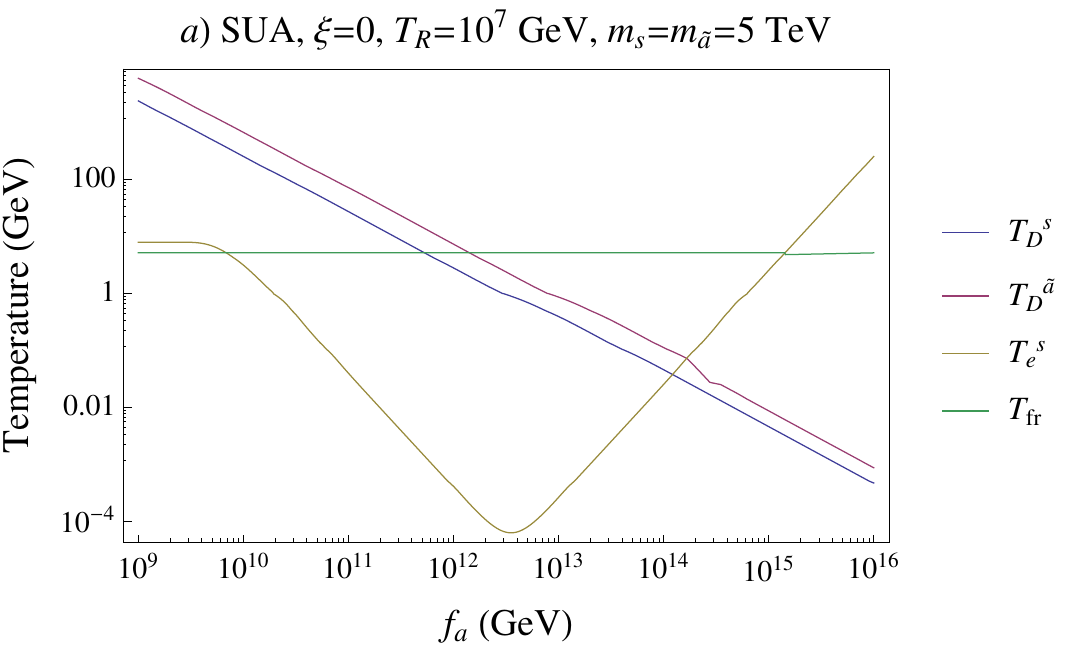}
\includegraphics[height=6.9cm]{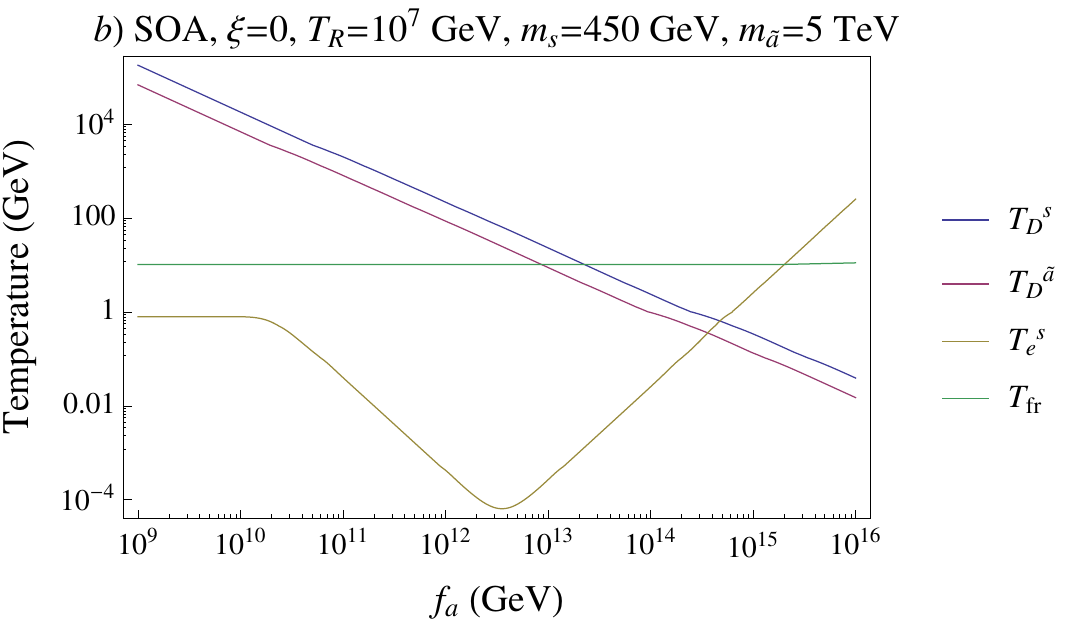}
\caption{Saxion (blue) and axino (purple) decay temperature along with saxion-radiation equality temperature (yellow) are shown for
{\it a}) the SUA and {\it b}) the SOA benchmark point for $\xi =0$.
The neutralino freeze-out temperature is also shown.
We use $T_R=10^7$ GeV and $m_s=m_{\ta}=5$ TeV for {\it a}) and $m_s=450$ GeV, $m_{\ta}=5$ TeV for {\it b}).
\label{fig:sax_temp_xi=0}}
\end{center}
\end{figure}

For illustration, the various temperatures are shown as a function of $f_a$ for $\xi=0$
in Fig.~\ref{fig:sax_temp_xi=0}
for {\it a}) the SUA benchmark and in {\it b}) for the SOA benchmark.
We take $m_s=m_{\ta}=5$ TeV for SUA and $m_s=450$ GeV, $m_{\ta}=5$ TeV for SOA.
For low $f_a\lesssim 10^{12-13}$ GeV, saxions and axinos decay before the neutralino freeze-out
when the universe is radiation-dominated in which case the neutralino dark matter density is
determined by the usual freeze-out mechanism.
For $10^{12}$ GeV$\lesssim f_a\lesssim10^{14}$ GeV, saxions and axinos decay after neutralino freeze-out
so that the neutralino re-annihilation process  becomes important to determine the WIMP portion of the dark matter density.
For $f_a\gtrsim10^{14}$ GeV, saxion coherent oscillation can dominate over radiation and 
the saxion decay  occurs after neutralino freeze-out.
In this case, the WIMP abundance may be depleted by late-time entropy injection, 
or augmented if saxions decay at a large rate into SUSY particles.
%

\begin{figure}
\begin{center}
\includegraphics[height=6.9cm]{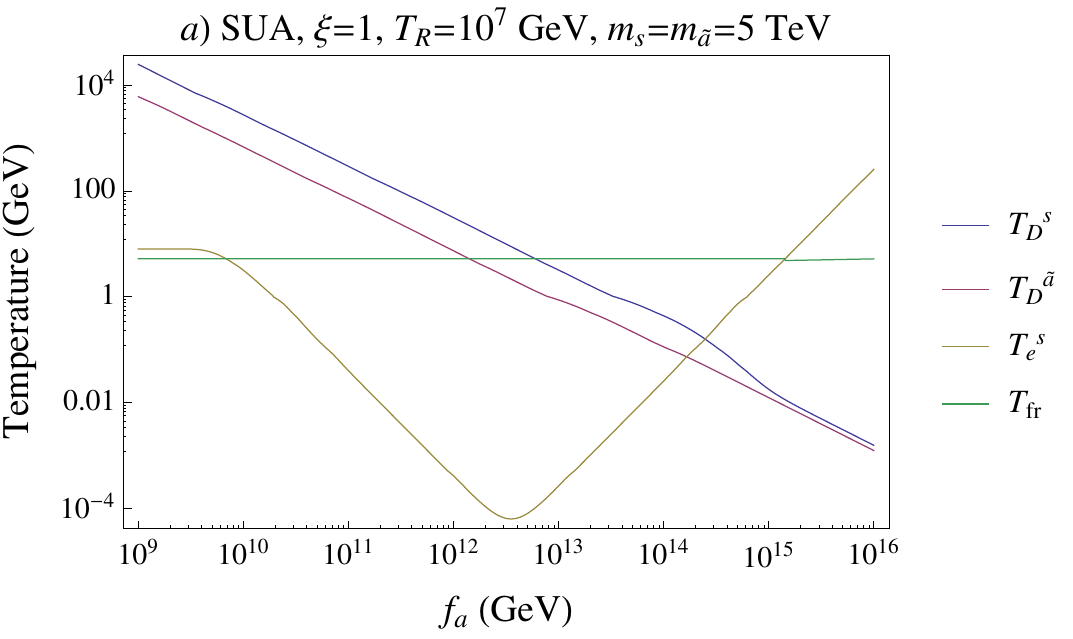}
\includegraphics[height=6.9cm]{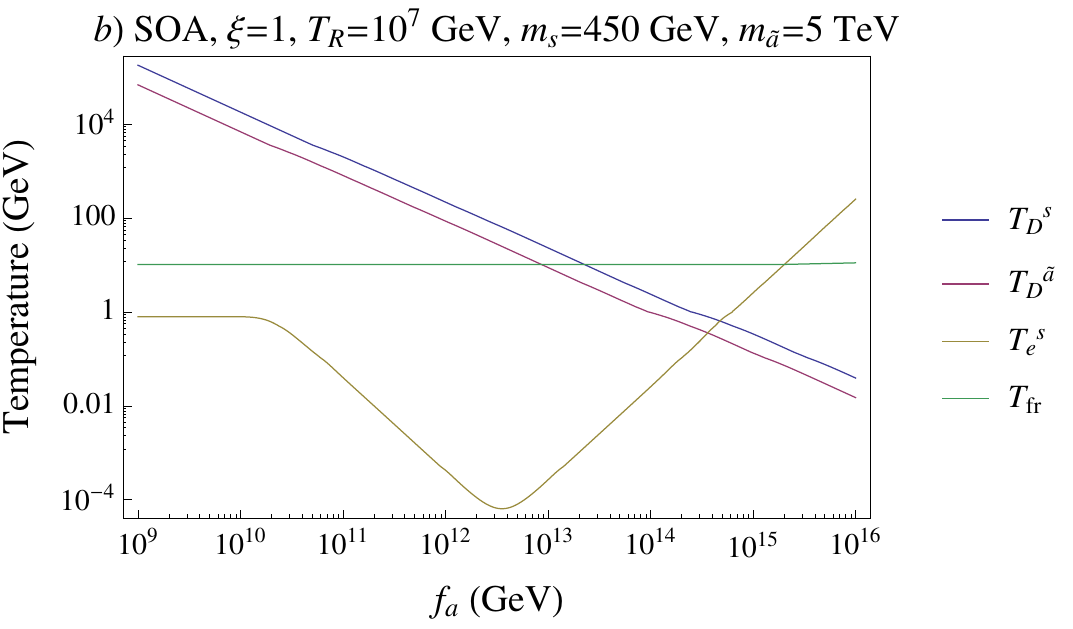}
\caption{Saxion (blue) and axino (purple) decay temperature along with saxion-radiation equality temperature (yellow) are shown for
{\it a}) the SUA and {\it b}) the SOA benchmark point for $\xi =1$.
The neutralino freeze-out temperature is also shown.
We use $T_R=10^7$ GeV and $m_s=m_{\ta}=5$ TeV for {\it a}) and $m_s=450$ GeV, $m_{\ta}=5$ TeV for {\it b}).
\label{fig:sax_temp_xi=1}}
\end{center}
\end{figure}

The case for $\xi =1$ is shown in Fig. \ref{fig:sax_temp_xi=1}.
For $\xi=1$, the decay $s\to aa$ (and possibly $s\to\ta\ta$) is allowed, which leads to an even earlier
saxion decay, but also to the possible production of dark radiation.
Since the saxion decay temperature is even higher than the $\xi=0$ case,
the equality occurs when $f_a$ is a few times larger than the $\xi=0$ case.

We are now ready to make a detailed study of several cosmological scenarios of the SUSY DSFZ model.

\section{Cosmological scenarios depending on $f_a$}
\label{sec:cosmo}

In this section, we will discuss various cosmological scenarios which have different characteristics depending on the
PQ scale $f_a$ ($=\sqrt{2}v_{PQ}$).  Our analysis will be presented separately for SUA and SOA.

\subsection{SUA}

As was discussed previously with Figs.~\ref{fig:sax_temp_xi=0}{\it a}) and \ref{fig:sax_temp_xi=1}{\it a}),
there are three regions of $f_a$ having different cosmological properties in terms of the dark matter abundance:
1.\ $f_a\lesssim10^{12}$ GeV, 2.\ $10^{12}$ GeV$\lesssim f_a\lesssim10^{14}$ GeV  and 3.\ $f_a\gtrsim 10^{14}$ GeV
for which saxions (axinos) decay 1.\ before dark matter freeze-out 2.\ after freeze-out and 3.\
dominate the universe before their decay.

\subsubsection{$f_a\lesssim10^{12}$ GeV}

In this region, axinos and saxions are produced mainly by thermal scattering and thus they do not dominate the universe.
Furthermore, they decay  before neutralino freeze-out so that the standard thermal relic density
$\Omega^{\rm std}_{\tilde Z_1} h^2 = 0.01$ remains valid.
In this region, then, the main component of dark matter would come from misalignment-produced cold axions.
The initial misalignment angle $\theta_i$ can always be adjusted so that $\Omega_ah^2 =0.12-\Omega_{\tz_1}h^2$.
Thus, for $f_a\alt 10^{12}$ GeV, we would expect in the SUA benchmark case a universe with dark matter at
$\sim 10\%$ Higgsino-like WIMPs along with 90\% cold axions~\cite{prl}.

\begin{figure}
\begin{center}
\includegraphics[height=6.9cm]{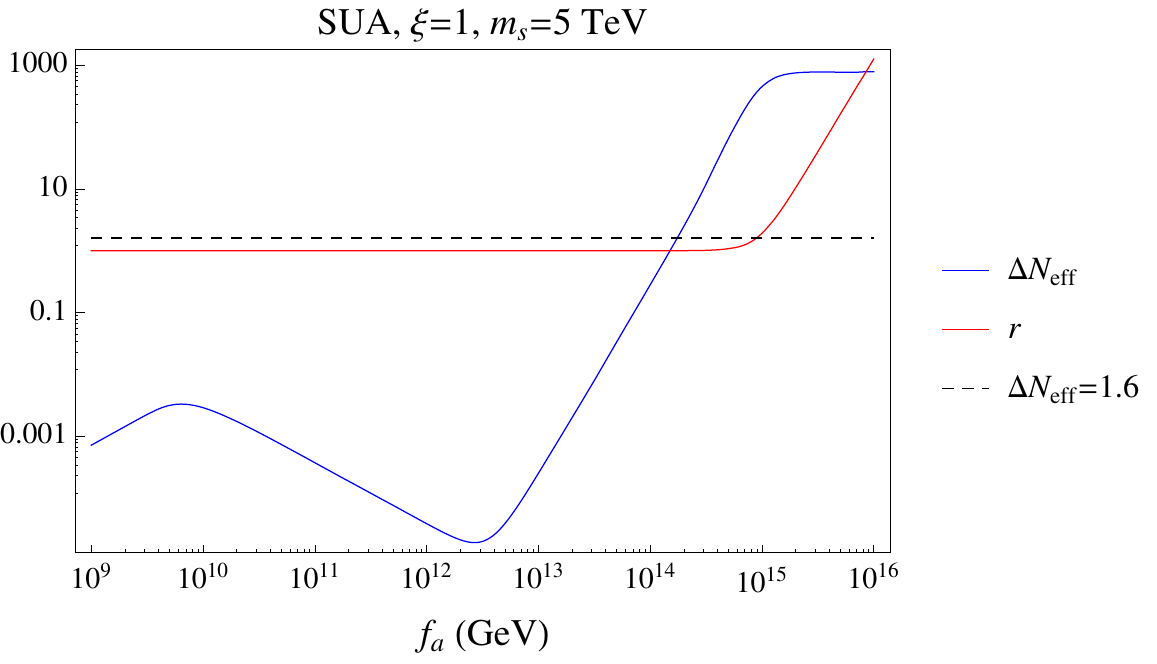}
\caption{Plot for $\Delta N_{\rm eff}$ and $r$ in SUA, $\xi=1$ and $m_s=5$ TeV.
\label{fig:DNeff_SUA_ms=5000}}
\end{center}
\end{figure}

It remains to check how sizable is the relativistic axion production from the saxion decay or thermal scattering.
The effective number of neutrinos from relativistic axions is given by
\begin{equation}
\Delta N_{\rm eff}\simeq\Delta N_{\rm eff}^{\rm TP}+\frac{18}{r}BR(s\to aa)g_{*S}(T_D)^{-1/3}
\frac{m_s\left(Y_s^{\rm CO}+Y_s^{\rm TP}\right)}{T_D^s}
\end{equation}
where $r$ is the factor of entropy dilution. In this region there is no entropy dilution ($r=1$).
In most cases, thermal production of axions is negligible and thus it is enough to only take the saxion decay into account.
The most important quantity is $BR(s\to aa)$ which was discussed in Sec.~\ref{sec:sdecay}.
For $m_s\gtrsim 4$ TeV, the dominant decay mode of the saxion is its decay to neutralinos and charginos.
Thus, we get to a good approximation,
\begin{eqnarray}
BR(s\to aa)&\approx& \frac{\Gamma(s\to aa)}{\Gamma\left(s\to \widetilde{Z}\widetilde{Z}\right)+\Gamma\left(s\to \widetilde{W}\widetilde{W}\right)+\Gamma(s\to aa)}\nonumber\\
&\approx&\frac{\xi^2m_s^3/(32\pi f_a^2)}{c_H^2\mu^2m_s/(16\pi f_a^2)+\xi^2m_s^2/(32\pi f_a^2)}\nonumber\\
&=&\frac{\xi^2m_s^2}{2c_H^2\mu^2+\xi^2m_s^2}
\end{eqnarray}
which leads to
\begin{eqnarray}
\Delta N_{\rm eff}&\simeq& 18\times g_{*S}(T_D)^{-1/3}\frac{\xi^2m_s^2}{2c_H^2\mu^2+\xi^2m_s^2}\frac{m_sY_s^{\rm TP}}{T_D^s}\nonumber\\
&\approx&6.6\times10^{-5}
\left(1+\frac{2c_H^2\mu^2}{\xi^2m_s^2}\right)^{-3/2}
\left(\frac{\zeta_s}{\xi}\right)
\left(\frac{g_{*S}}{90}\right)^{-1/3}
\left(\frac{g_*}{90}\right)^{1/4}\nonumber\\
&&\times\left(\frac{5\mbox{ TeV}}{m_s}\right)^{1/2}
\left(\frac{B\mu/\mu\mbox{ or }\mu}{\rm TeV}\right)^2
\left(\frac{10^{12}\mbox{ GeV}}{f_a}\right).
\label{eq:DNeff_low_fa}
\end{eqnarray}
In the range of $10^{10}$ GeV$\lesssim f_a\lesssim 10^{12}$ GeV, one finds $\Delta N_{\rm eff}$ typically
between $10^{-3}$ and $10^{-5}$ even with $\xi =1$ as can be seen clearly from Fig.~\ref{fig:DNeff_SUA_ms=5000}.
Note also that the saxion density approaches its equilibrium value for $f_a\lesssim 10^{10}$ GeV so that
$\Delta N_{\rm eff}$ does not exceed $10^{-3}$ even for smaller $f_a$.
Therefore, we can conclude that the relativistic axion abundance is far below the current limit on dark radiation from PLANCK~\cite{planck}.

\subsubsection{$10^{12}$ GeV$\lesssim f_a\lesssim10^{14}$ GeV}

\begin{figure}
\begin{center}
\includegraphics[height=6.9cm]{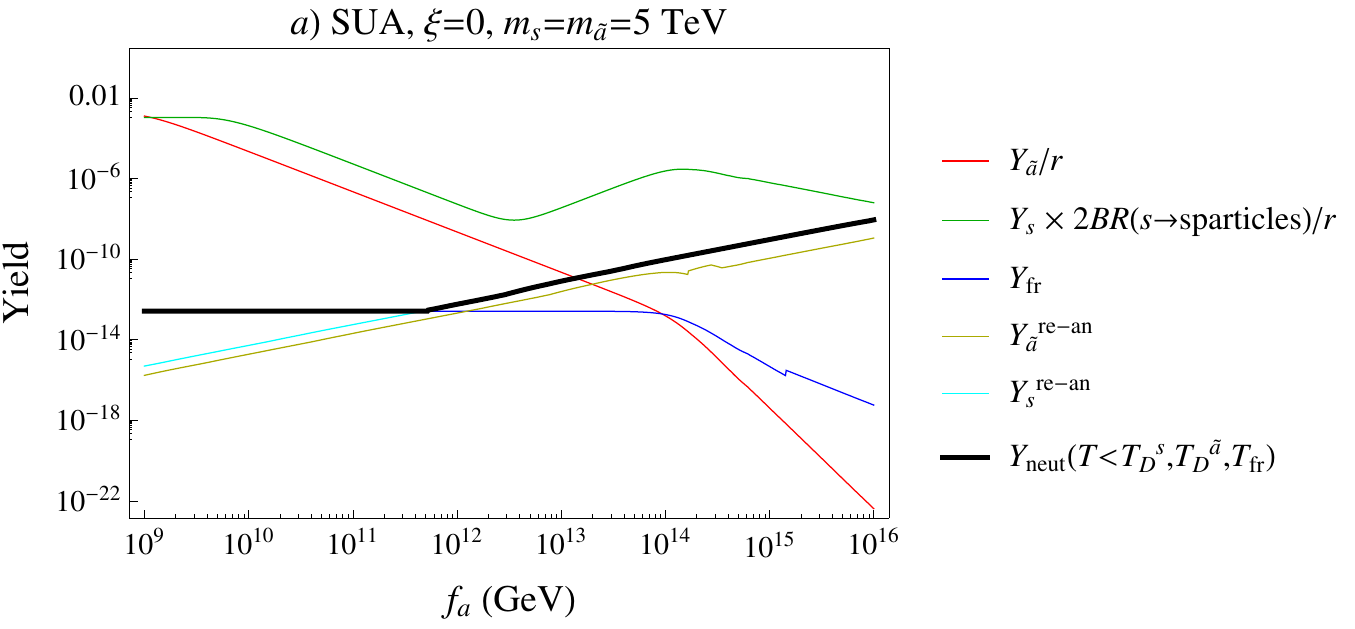}
\includegraphics[height=6.9cm]{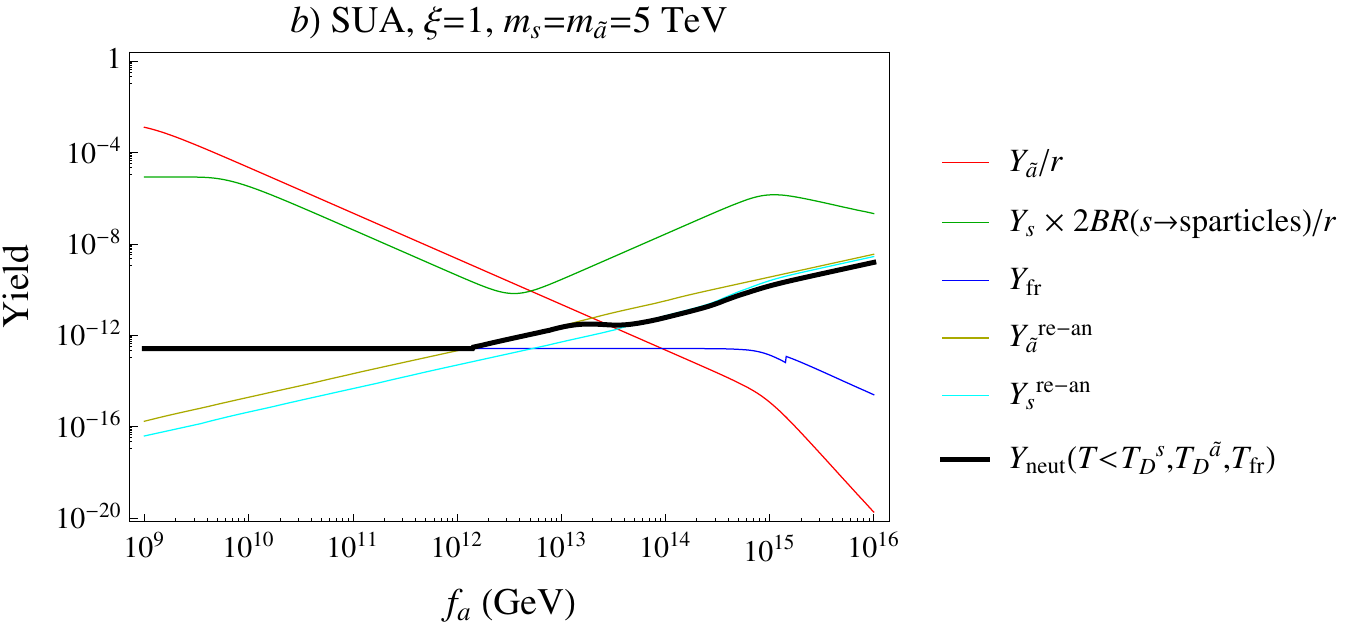}
\caption{Neutralino yield from standard freeze-out and decay. We use {\it a}) $\xi=0$ and {\it b}) $\xi=1$.
$Y_{\ta}^{\rm re-an}$ ($Y_{s}^{\rm re-an}$) is the re-annihilation contribution for neutralino density from axino (saxion) decay. $Y_{\rm neut}$ is the final neutralino yield.
\label{fig:neut_rean_SUA_ms=5000}}
\end{center}
\end{figure}

In this region, saxions and axinos do not dominate the universe, but they do decay after neutralino freeze-out.
In addition, saxion production from coherent oscillation becomes larger than thermal production
as shown in Figs.~\ref{fig:sax_temp_xi=0}{\it a}) and \ref{fig:sax_temp_xi=1}{\it a}).
Such late decays of saxions and axinos can produce an overabundance of neutralino dark matter particles
which then re-annihilate to deplete their initial density.
To discuss the neutralino dark matter density in this region, let us consider the saxion decay temperature:
\begin{equation}
T_D^s\simeq 3.1\mbox{ GeV}\times \xi\left(1+\frac{2c_H^2\mu^2}{\xi^2m_s^2}\right)^{1/2}
\left(\frac{90}{g_*}\right)^{1/4}\left(\frac{m_s}{5\mbox{ TeV}}\right)^{3/2}
\left(\frac{10^{13}\mbox{ GeV}}{f_a}\right).
\end{equation}
Note that the standard neutralino freeze-out temperature $T_{\rm fr}$ is around 5 GeV.
Thus, the saxion decay temperature gets smaller than $T_{\rm fr}$ when $f_a\gtrsim 10^{13}$ GeV for $\xi\sim1$,
or  when $f_a\gtrsim10^{12}$ GeV for $\xi\lesssim0.1$.
The axino decay temperature is
\begin{equation}
T_D^{\ta}\simeq 3.7\mbox{ GeV}\times c_H\left(\frac{90}{g_*}\right)^{1/4}
\left(\frac{\mu}{150\mbox{ GeV}}\right)
\left(\frac{10^{12}\mbox{ GeV}}{f_a}\right)
\left(\frac{m_{\ta}}{5\mbox{ TeV}}\right)^{1/2},
\end{equation}
which can be smaller  $T_{\rm fr}$ when $f_a\gtrsim 10^{12}$ GeV.
The neutralino density determined from the re-annihilation process is given by
\begin{eqnarray}
Y^{-1}_{\widetilde{Z}_1}(T<T_D)&\simeq& Y_{\widetilde{Z}_1}^{-1}(T_D)+
\left(Y_{\widetilde{Z}_1}^{\rm re-an}\right)^{-1}\nonumber\\
&=&Y_{\widetilde{Z}_1}^{-1}(T_D)+
\frac{4\langle \sigma v\rangle M_pT_D}{\left(90/\pi^2g_*(T_D)\right)^{1/2}}
\label{eq:neut_rean}
\end{eqnarray}
where $Y_{\widetilde{Z}_1}(T_D)=Y_{\widetilde{Z}_1}^{\rm fr}+Y_{\widetilde{Z}_1}^{\rm decay}$ and $T_D$ can be either $T_D^s$ or $T_D^{\ta}$.
For $10^{12}$ GeV$\lesssim f_a\lesssim10^{14}$ GeV, neutralino production due to axino/saxion decay is
then much larger than that from the standard neutralino freeze-out, {\it i.e.}
$Y_{\widetilde{Z}_1}(T_D) \simeq Y_{\widetilde{Z}_1}^{\rm decay}$.
The neutralino abundance dominated by the re-annihilation term $Y_{\widetilde{Z}_1}^{\rm re-an}$ of Eq.~(\ref{eq:neut_rean})
is approximated by
\begin{eqnarray}
Y_{\widetilde{Z}_1}(T<T_D) &\approx&
4.1\times10^{-13}\frac{1}{c_H}\left(\frac{g_*}{90}\right)^{1/4}
\left(\frac{2.57\times10^{-25}\mbox{ cm}^3/\mbox{s}}{\langle \sigma v\rangle}\right)\nonumber\\
&&\times\left(\frac{150\mbox{ GeV}}{\mu}\right)
\left(\frac{f_a}{10^{12}\mbox{ GeV}}\right)
\left(\frac{5\mbox{ TeV}}{m_{\ta}}\right)^{1/2}.
\label{eq:neut_after_rean}
\end{eqnarray}
The schematic behavior  of the neutralino yield from standard freeze-out, axino decay, saxion decay and re-annihilation
is shown in Fig.~\ref{fig:neut_rean_SUA_ms=5000}.

One of the most important features is that the neutralino density can be larger than the standard density for $f_a\gtrsim 10^{12}$ GeV.
This can cause a conflict with  the direct detection bound from XENON100 since Higgsino-like WIMPs have a large spin-independent
nucleon scattering cross-section.
From the neutralino yield via re-annihilation, Eq.~(\ref{eq:neut_after_rean}),
we get the neutralino dark matter density
\begin{eqnarray}
\Omega_{\widetilde{Z}_1}h^2&\simeq& \frac{m_{\widetilde{Z}_1} Y_{\widetilde{Z}_1}}{3.6\, \mbox{eV}}
\nonumber\\
&=&0.015\times{c_H}\left(\frac{g_*}{90}\right)^{1/4}
\left(\frac{2.57\times10^{-25}\mbox{ cm}^3/\mbox{s}}{\langle \sigma v\rangle}\right)\nonumber\\
&&\times\left(\frac{150\mbox{ GeV}}{\mu}\right)
\left(\frac{f_a}{10^{12}\mbox{ GeV}}\right)
\left(\frac{5\mbox{ TeV}}{m_{\ta}}\right)^{1/2}
\left(\frac{m_{\widetilde{Z}_1}}{135\mbox{ GeV}}\right).
\end{eqnarray}
This has to be compared with  the neutralino density bound from the XENON100 experiment~\cite{xenon}
for the SUA benchmark point which is
\begin{equation}
\Omega_{\widetilde{Z}_1}^{Xe}h^2<0.026.
\end{equation}
Therefore, we get the bound: $f_a/c_H \lesssim 2 \times 10^{12}$ GeV, which again requires
the cold axion as a major component of dark matter.  Of course, this constraint can be avoided for the case of
a more purely Higgsino-like dark matter scenario (with larger bino/wino mass) to saturate the dark matter density
($\Omega_{\tilde Z_1} h^2=0.11$)  by the re-annihilation process.
That is, we can open the possibility for Higgsino-like dark matter which is more abundant than in the standard cosmology
in this region of $f_a$.

\medskip

Concerning the constraint from dark radiation, the situation is not much different from the case of $f\lesssim10^{12}$ GeV.
Although saxion and axino decay after neutralino freeze-out, there is no matter domination era in this region, and thus
the produced axion abundance from saxion decay is described again by Eq.~(\ref{eq:DNeff_low_fa}) which gives an
even smaller amount of dark radiation for larger $f_a$.

\subsubsection{$f_a\gtrsim10^{14}$ GeV}

Although thermally produced saxions and axinos do not dominate the universe at large $f_a$,
oscillation production of saxions can dominate for large enough $f_a$  as discussed in the previous section.
We rewrite the condition for coherent oscillations of saxions to dominate the universe:
\begin{equation}
f_a\gtrsim 2.7\times10^{14}\mbox{ GeV}\times \xi^{1/3}
\left(1+\frac{2c_H^2\mu^2}{\xi^2m_s^2}\right)^{1/6}
\left(\frac{10}{g_*}\right)^{1/12}
\left(\frac{10^7\mbox{ GeV}}{{\rm min}[T_R,T_s]}\right)^{1/3}
\left(\frac{m_s}{\rm 5 TeV}\right)^{1/2}.
\end{equation}
In this region, cosmology becomes more interesting.
As the saxion coherent oscillation dominates the universe, the neutralino dark matter density,
radiation and dark radiation are all determined
by branching ratios of the saxion decays into sparticles, SM particles and axions.

\begin{figure}
\begin{center}
\includegraphics[height=5.9cm]{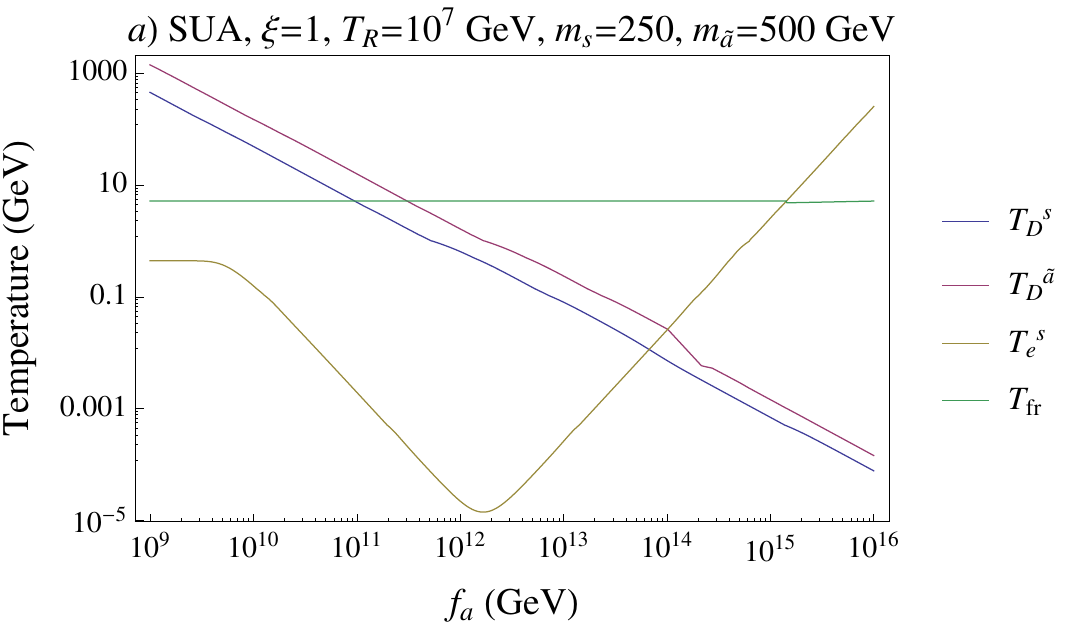}~~~~~~~~~~~~~~~~~~~~~\\
\includegraphics[height=5.9cm]{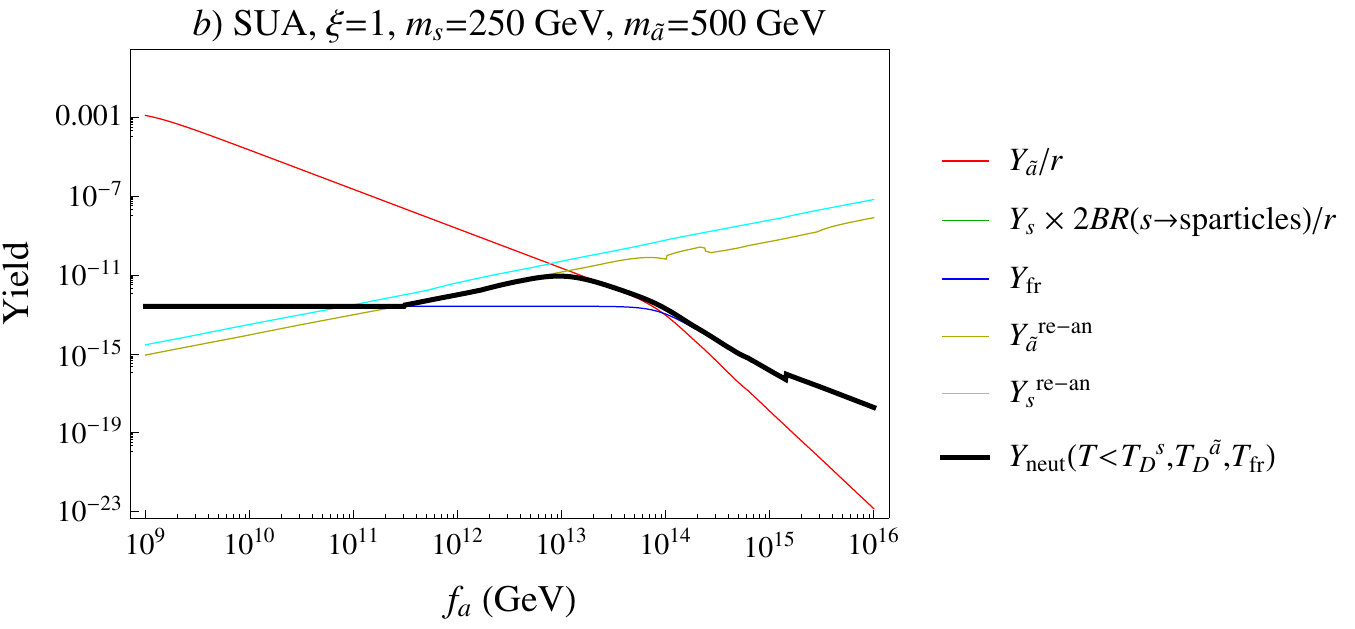}
\includegraphics[height=5.9cm]{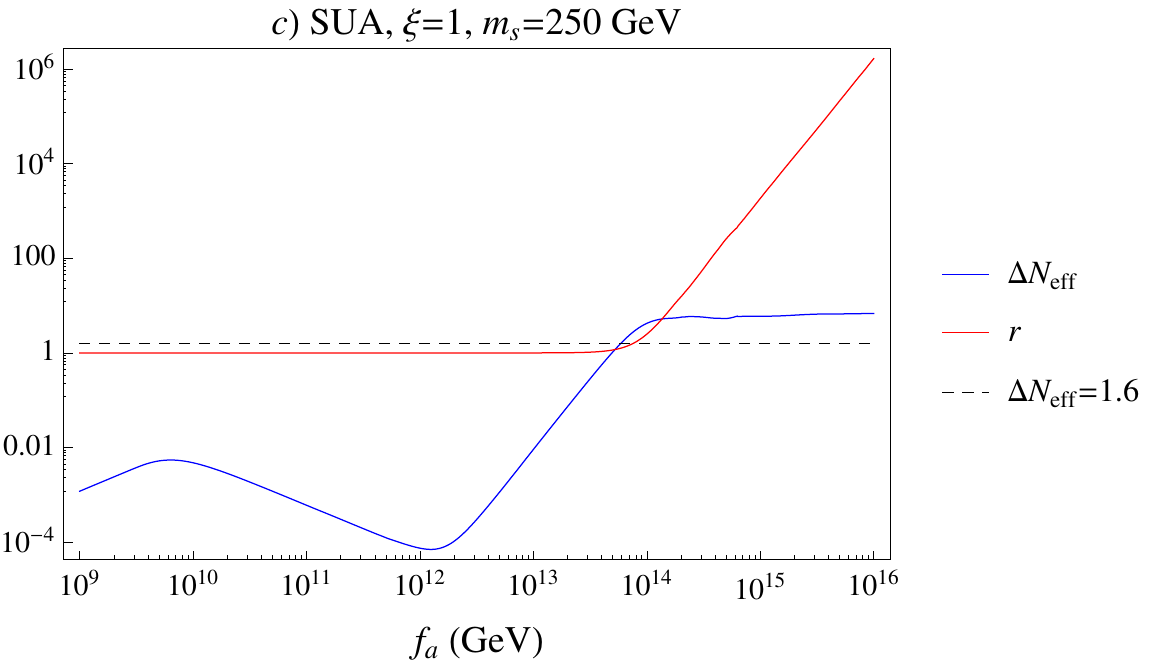}~~~~~~~~~~~
\caption{Plots for the SUA case with $\xi=1$, $T_R=10^7$ GeV and $m_s=250$ GeV.
{\it a}) Saxion decay temperature (blue), saxion-radiation equality temperature (yellow) and
neutralino freeze-out temperature in the standard cosmology (green) are shown.
{\it b}) Neutralino yield from standard freeze-out and decay.
{\it c}) $\Delta N_{\rm eff}$ and $r$.
\label{fig:neut_rean_SUA_xi=1_ms=250}}
\end{center}
\end{figure}

Under the  sudden decay approximation, we can obtain the number of effective neutrinos~\cite{choi97}:
\begin{eqnarray}
\Delta N_{\rm eff}&=&18\times\frac34g_*(T_D^s)^{-1/3}
\left[\frac{BR(s\to aa)}{1-BR(s\to aa)}\right]\nonumber\\
&\simeq&14\times g_*(T_D^s)^{-1/3}\frac{\xi^2m_s^2}{2c_H^2\mu^2}\nonumber\\
&=&3.5\times10^3~
\frac{\xi^2}{c_H^2}\left(\frac{10}{g_*(T_D^s)}\right)^{1/3}
\left(\frac{m_s}{5\mbox{ TeV}}\right)^2
\left(\frac{150\mbox{ GeV}}{\mu}\right)^2.
\label{eq:DNeff_SUA}
\end{eqnarray}
Note that we assume that most of the neutralinos produced by saxion decay re-annihilate into SM particles
which eventually contribute to radiation.
The formula shows that the dark radiation constraint is very severe in this region of large $f_a$.
This arises from the fact that saxion decay into axion pairs is the dominant mode for large $m_s$.
If the saxion mass is around $\mu=150$ GeV,  it is possible to obtain $\Delta N_{\rm eff}\lesssim1$
(see an example in Fig.~\ref{fig:neut_rean_SUA_xi=1_ms=250}{\it c})).
Otherwise, $\xi$ should be suppressed to be ${\cal O}(0.01)$.
Note that a smaller saxion mass makes the saxion decay temperature smaller down to ${\cal O}(1)$ MeV
for $f_a\gtrsim10^{14}$ GeV (see Eq.~(\ref{eq:sax_TD_high_fa})) so that such a region is now constrained by BBN
(see Fig.~\ref{fig:neut_rean_SUA_xi=1_ms=250}{\it a})).

In Fig.~\ref{fig:DNeff_SUA_ms=5000}, we show the $\Delta N_{\rm eff}$ for $m_s=5$ TeV.
As discussed before, $\Delta N_{\rm eff}$ is well below the current limit for $f_a\lesssim 10^{14}$ GeV.
For $f_a\gtrsim 10^{14}$ GeV, $\Delta N_{\rm eff}$ become larger than the current limit of $1.6$, 
so this parameter region of $f_a$ becomes excluded.
Let us note that a terminal value of $\Delta N_{\rm eff}$ described by Eq.~(\ref{eq:DNeff_SUA}) 
is reached for $f_a\gtrsim 10^{15}$ GeV.
This is due to the fact that {\it only part of saxion decay contributes to the radiation energy 
so that the entropy dilution takes place for rather larger value of $f_a$ than that of saxion domination.}
The saxion domination and entropy dilution will be discussed in the following paragraphs.

After the period of the saxion domination, its decay overproduces neutralinos and their relic density is again
determined by re-annihilation,
\begin{equation}
Y_{\widetilde{Z}_1}^{-1}\simeq
\left(Y_{\widetilde{Z}_1}^{\rm fr}+Y_{\widetilde{Z}_1}^{\rm decay}\right)^{-1}\times r
+\frac{4\langle \sigma v\rangle M_pT^s_D}{\left(90/\pi^2g_*(T^s_D)\right)^{1/2}},
\label{eq:neut_den_sax_dom}
\end{equation}
where $r$ is the entropy dilution factor which is given by
\begin{equation}
r=\mbox{max}\left[1,\frac43\left[1-BR(s\to aa)\right]\frac{Y_sm_s}{T_D^s}\right].
\end{equation}
The decay temperature of the saxion is determined by the visible energy density from the saxion decay
and is given by~\cite{Jeong12}
\begin{eqnarray}
T_D^s&=&\left[\left\{1-BR(s\to aa)\right\}\frac{90}{\pi^2g_*}\right]^{1/4}\sqrt{\Gamma_s M_p}\nonumber\\
&\simeq&0.11\mbox{ GeV}\sqrt{\xi c_H}
\left(1+\frac{2c_H^2\mu^2}{\xi^2m_s^2}\right)^{1/4}
\left(\frac{10}{g_*}\right)^{1/4}\nonumber\\
&&\times
\left(\frac{\mu}{150\mbox{ GeV}}\right)^{1/2}
\left(\frac{m_s}{5\mbox{ TeV}}\right)
\left(\frac{10^{14}\mbox{ GeV}}{f_a}\right).
\label{eq:sax_TD_high_fa}
\end{eqnarray}
In this parameter region, neutralinos are produced mostly by saxion decay,
and thus the first term of Eq.~(\ref{eq:neut_den_sax_dom}) is simply given by
\begin{eqnarray}
Y_{\widetilde{Z}_1}^{\rm decay}&\simeq& Y_s^{\rm CO}\times 2 BR(s\to\mbox{sparticles})\nonumber\\
&\simeq&6.8\times 10^{-5}~\frac{c_H^2}{\xi^2}
\left(1+\frac{2c_H^2\mu^2}{\xi^2m_s^2}\right)^{-1}
\left(\frac{\mbox{min}[T_R,T_s]}{10^7\mbox{ GeV}}\right)\nonumber\\
&&\times\left(\frac{f_a}{10^{14}\mbox{ GeV}}\right)^2
\left(\frac{\mu}{150\mbox{ GeV}}\right)^2
\left(\frac{5\mbox{ TeV}}{m_s}\right)^3
\end{eqnarray}
if there is no dilution, or
\begin{eqnarray}
Y_{\widetilde{Z}_1}^{\rm decay}\times\frac1r\simeq
\frac{3Y_s\times 2BR(s\to\mbox{sparticles})T_D^s}{4\left[1-BR(s\to aa)\right]Y_sm_s}
\end{eqnarray}
if the dilution factor $r$ is larger than unity.
For large enough $m_s$, we have $1-BR(s\to aa)\simeq BR(s\to\mbox{sparticles})$ leading to
\begin{eqnarray}
Y_{\widetilde{Z}_1}^{\rm decay}\times\frac1r\simeq
\frac{3T_D^s}{2m_s}&=&3.3\times10^{-5}\sqrt{\xi c_H}
\left(1+\frac{2c_H^2\mu^2}{\xi^2m_s^2}\right)^{1/4}
\left(\frac{10}{g_*}\right)^{1/4}\nonumber\\
&&\times
\left(\frac{\mu}{150\mbox{ GeV}}\right)^{1/2}
\left(\frac{10^{14}\mbox{ GeV}}{f_a}\right).
\end{eqnarray}
Now, the re-annihilation part becomes
\begin{eqnarray}
Y_{\widetilde{Z}_1}^{\rm re-an}&\equiv&
\frac{\left(90/\pi^2g_*(T^s_D)\right)^{1/2}}{4\langle \sigma v\rangle M_pT^s_D}\nonumber\\
&\simeq&4.1\times10^{-11}~\frac{1}{\sqrt{\xi c_H}}
\left(\frac{10}{g_*}\right)^{1/4}
\left(1+\frac{2c_H^2\mu^2}{\xi^2m_s^2}\right)^{-1/4}
\left(\frac{2.57\times10^{-25}\mbox{ cm}^3/\mbox{s}}{\langle \sigma v\rangle}\right)\nonumber\\
&&\times\left(\frac{150\mbox{ GeV}}{\mu}\right)^{1/2}
\left(\frac{f_a}{10^{14}\mbox{ GeV}}\right)
\left(\frac{5\mbox{ TeV}}{m_s}\right).
\end{eqnarray}
Comparing $Y_{\widetilde{Z}_1}^{\rm decay}/r$ and $Y_{\widetilde{Z}_1}^{\rm re-an}$,
we find that the re-annihilation dominantly determines the neutralino density for $f_a\lesssim 10^{17}$ GeV.
Therefore, the abundance of neutralinos is
\begin{eqnarray}
\Omega_{\widetilde{Z}_1}h^2&=&
\left(\frac{m_{\widetilde{Z}_1} Y_{\widetilde{Z}_1} }{ 3.6\, \rm eV}\right)
\nonumber\\
&\simeq& 1.5~\frac{1}{\sqrt{\xi c_H}}
\left(\frac{10}{g_*}\right)^{1/4}
\left(1+\frac{2c_H^2\mu^2}{\xi^2m_s^2}\right)^{-1/4}
\left(\frac{2.57\times10^{-25}\mbox{ cm}^3/\mbox{s}}{\langle \sigma v\rangle}\right)\nonumber\\
&&\times
\left(\frac{150\mbox{ GeV}}{\mu}\right)^{1/2}
\left(\frac{f_a}{10^{14}\mbox{ GeV}}\right)
\left(\frac{5\mbox{ TeV}}{m_s}\right)
\left(\frac{m_{\widetilde{Z}_1}}{135\,\rm GeV}\right).
\end{eqnarray}
The schematic plots for neutralino yield are shown in Fig.~\ref{fig:neut_rean_SUA_ms=5000}.
The neutralino abundance is much larger than the observed value for $f_a=10^{14}$ GeV.
This provides a serious constraint for the model together with $\Delta N_{\rm eff}$.

To avoid the problem of overclosure density of neutralinos,
we may consider a case with a light enough saxion such that decay to neutralinos is forbidden;
then saxion decay produces only axion pairs and entropy.
The various temperatures, yields and $\Delta N_{eff}$ are shown in frames {\it a}), {\it b}) and {\it c}) of
Fig.~\ref{fig:neut_rean_SUA_xi=1_ms=250}.
In this case, the existing relic particles are diluted away  as shown in frame {\it b}).
Even in this case, however, BBN strongly constrains the large $f_a$ region as discussed previously.

\subsection{SOA}

Similarly to the SUA case, we divide the region of $f_a$ into three parts:
$f_a\lesssim 10^{13}$ GeV, $10^{13}$ GeV$\lesssim f_a\lesssim10^{14}$ and $f_a\gtrsim 10^{14}$ GeV, which correspond to the
regions of the saxion/axino decay before the neutralino freeze-out, after the freeze-out,
and the saxion domination before its decay, respectively.
One crucial difference arises due to the fact that $\mu$ is very large for SOA compared to the SUA case.
Such a large $\mu$ makes the saxion decay temperature (for $m_s\lesssim 5$ TeV)  one or two orders of magnitude larger than
the SUA case as shown in Figs.~\ref{fig:sax_TD_xi=0}{\it a}) and \ref{fig:sax_TD_xi=1}{\it b}).
As discussed in the SUA case, large $f_a$ might cause overproduction of neutralinos and relativistic axions from the saxion decay.
Such problems can be avoided by considering a light saxion that does not decay into sparticle pairs, and a small $\xi$ to suppress
$BR(s\to aa)$. However, the conflict with BBN coming from the saxion decay temperature close to ${\cal O}(1)$ MeV
is hardly circumvented  as shown  in Eq.~(\ref{eq:sax_TD_high_fa}).
In the SOA case, this tension is relieved as
the saxion decay temperature is enhanced by large $\mu$ even for lighter saxion masses.
Thus, we will take a smaller saxion mass to discuss cosmological implications for the SOA benchmark point.

\begin{figure}
\begin{center}
\includegraphics[height=6.9cm]{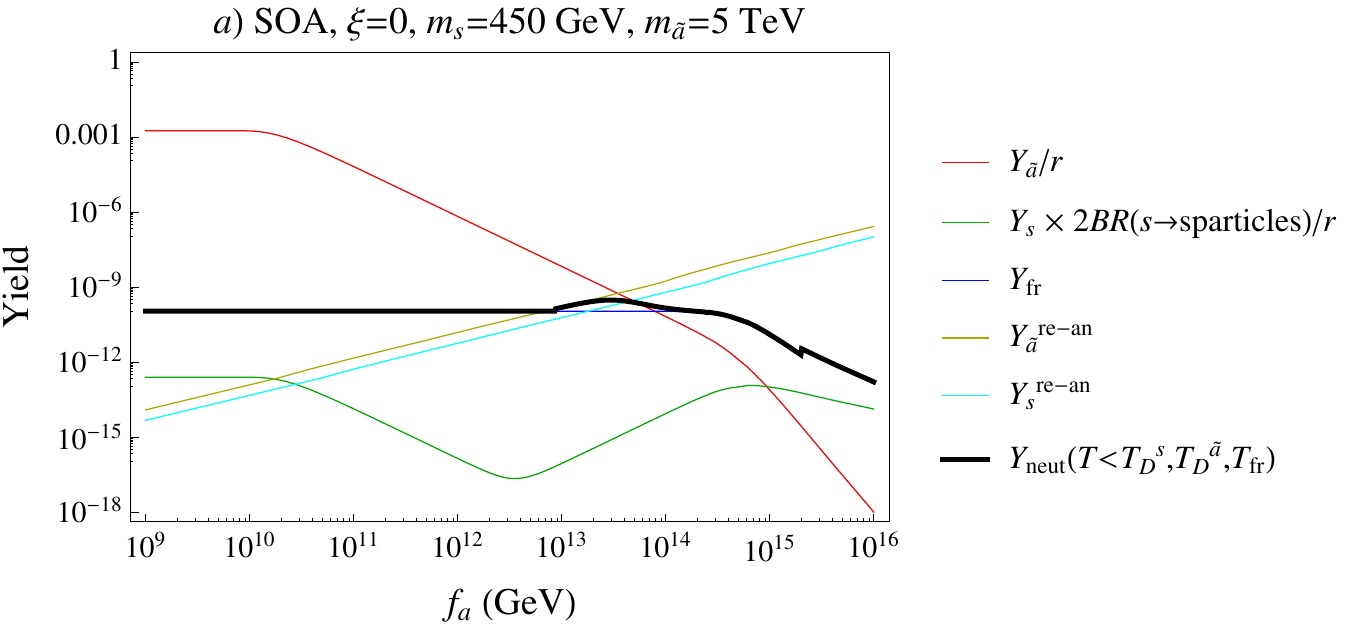}
\includegraphics[height=6.9cm]{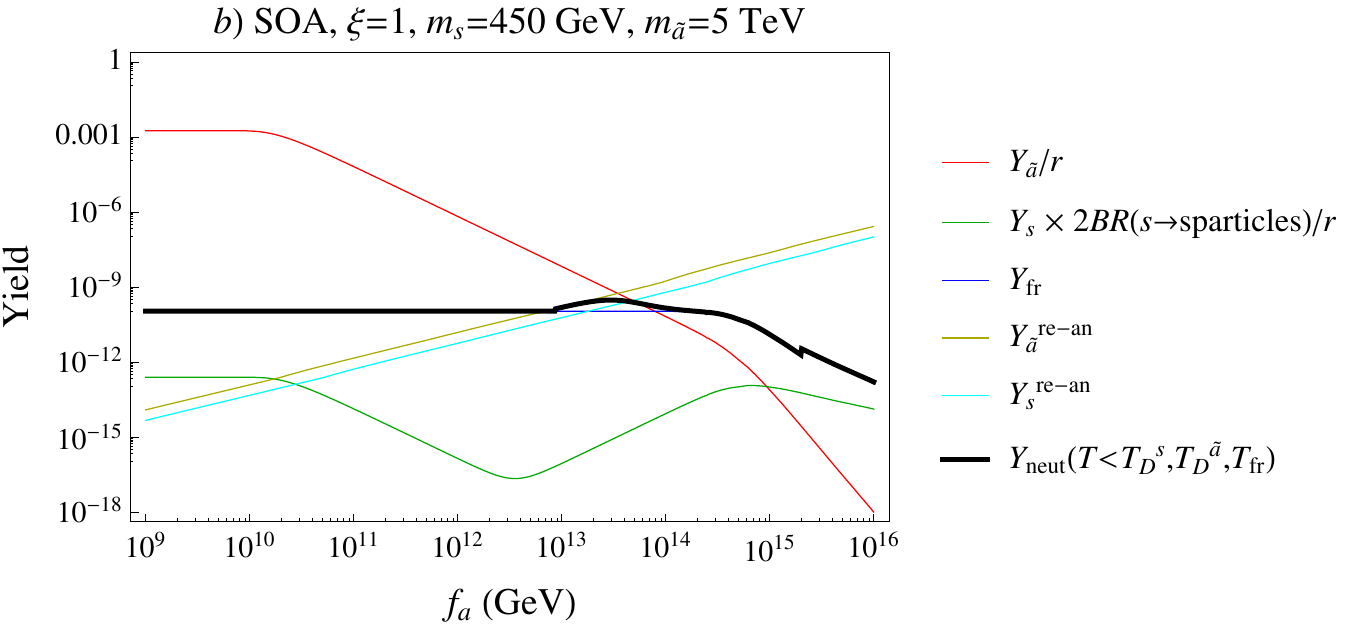}
\caption{Neutralino yield from standard freeze-out and decay. We use {\it a}) $\xi=0$ and {\it b}) $\xi=1$.
\label{fig:neut_rean_SOA}}
\end{center}
\end{figure}

\subsubsection{$f_a\lesssim 10^{13}$ GeV}

This region is basically ruled out by overproduction of the neutralino dark matter:
since saxions and axinos decay before neutralino freeze-out (see Figs.~\ref{fig:sax_temp_xi=0}{\it b}) and \ref{fig:sax_temp_xi=1}{\it b})),
the standard relic overabundance is unaltered: $\Omega_{\tilde Z_1} h^2=6.8$.

\begin{figure}
\begin{center}
\includegraphics[height=6.9cm]{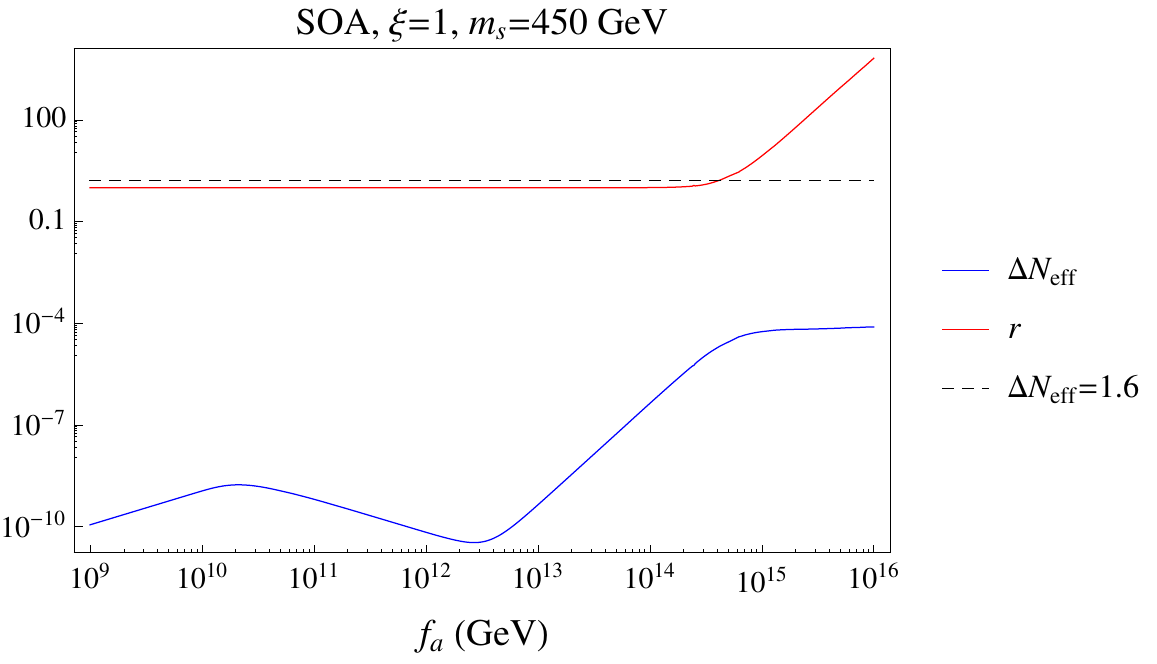}
\caption{Plot for $\Delta N_{\rm eff}$ and $r$ in SOA, $\xi=1$ and $m_s=450$ GeV.
\label{fig:DNeff_SOA}}
\end{center}
\end{figure}

For the effective number of neutrinos, we can rewrite Eq.~(\ref{eq:DNeff_low_fa}) inserting the SOA benchmark parameters:
\begin{eqnarray}
\Delta N_{\rm eff}&\approx&1.6\times10^{-10}~\frac{\xi^2\zeta_s}{c_H}
\left(1+\frac{\xi^2m_s^4}{16c_H^2\mu^4}\right)^{-3/2}
\left(\frac{g_{*s}}{90}\right)^{-1/3}
\left(\frac{g_*}{90}\right)^{1/4}\nonumber\\
&&\times\left(\frac{2.6\mbox{ TeV}}{\mu}\right)^6
\left(\frac{B\mu/\mu\mbox{ or }\mu}{\rm TeV}\right)^2
\left(\frac{10^{12}\mbox{ GeV}}{f_a}\right)
\left(\frac{m_s}{500\mbox{ GeV}}\right)^{11/2}.
\end{eqnarray}
Here we used the fact that the only relevant decay modes of the saxion are
$s\to hh$, $W^+W^-$, $ZZ$ and $aa$, which leads to the approximate expressions of $T_D^s$ and $BR(s\to aa)$
for $m_s\sim450$ GeV as follows:
\begin{eqnarray}
T_D^s&\approx&\left(\frac{c_H^2\mu^4}{4\pi v_{PQ}^2m_s}\right)^{1/2}
\left(1+\frac{\xi^2m_s^4}{16c_H^2\mu^4}\right)^{1/2}
M_p^{1/2}\left(\frac{90}{\pi^2g_*}\right)^{1/4},\\
BR(s\to aa)&\approx&
\frac{\xi^2m_s^4}{16c_H^2\mu^4}
\left(1+\frac{\xi^2m_s^4}{16c_H^2\mu^4}\right)^{-1}.
\end{eqnarray}
Again  $\Delta N_{\rm eff}$ is negligibly small (see Fig.~\ref{fig:DNeff_SOA}).

\subsubsection{$10^{13}$ GeV$\lesssim f_a\lesssim10^{14}$ GeV}

In this region, saxions and axinos decay after the neutralino freeze-out but still decay before
matter domination can take place.
The neutralino density can thus be augmented by saxion or axino decay while there is no entropy dilution.
Therefore, this region is also excluded by dark matter overproduction.

\subsubsection{$f_a\gtrsim 10^{14}$ GeV}

In this region, the saxion coherent oscillation can dominate the universe and thus the saxion decays into SM particles,
sparticles and axions determine the important cosmological quantities, {\it i.e.}, the amounts of entropy dilution,
neutralino density and dark radiation.

As in the SUA case (with $r>1$), the number of effective neutrinos is determined by
\begin{eqnarray}
\Delta N_{\rm eff}
&\simeq&14\times g_*(T_D^s)^{-1/3}\frac{\xi^2m_s^4}{16c_H^2\mu^4}\nonumber\\
&=&3.7\times10^{-4}~
\frac{\xi^2}{c_H^2}\left(\frac{10}{g_*(T_D^s)}\right)^{1/3}
\left(\frac{m_s}{450\mbox{ GeV}}\right)^4
\left(\frac{2.6\mbox{ TeV}}{\mu}\right)^4.
\label{eq:DNeff_SOA}
\end{eqnarray}
Thus, $\Delta N_{\rm eff}$ is negligible for the case of the SOA parameters.

The most important feature resides in the neutralino density.
The neutralino production from the saxion and axino decay is not very large as shown in Fig.~\ref{fig:neut_rean_SOA},
but there is a huge amount of entropy produced since the saxion dominantly decays
into Higgs and gauge boson states as can be seen in  Figs.~\ref{fig:sax_BR_xi=0}{\it b}) and \ref{fig:sax_BR_xi=1}{\it b}).
The dilution factor $r$ is given by
\begin{eqnarray}
r=\frac{T_e^s}{T_D^s}&\approx&
23\times c_H^{-1}\left(\frac{90}{g_*}\right)^{-1/4}
\left(1+\frac{\xi^2m_s^4}{16c_H^2\mu^4}\right)^{-1/2}\nonumber\\
&&\times\left(\frac{2.6\mbox{ TeV}}{\mu}\right)^2
\left(\frac{{\rm min}[T_R,T_s]}{10^7\mbox{ GeV}}\right)
\left(\frac{f_a}{10^{15}\mbox{ GeV}}\right)^3
\left(\frac{m_s}{450\mbox{ GeV}}\right)^{1/2} .
\end{eqnarray}
Here we used the saxion decay temperature given by
\begin{eqnarray}
T_D^s
&\approx&0.12\mbox{ GeV}\times c_H\left(\frac{90}{g_*}\right)^{1/4}
\left(1+\frac{\xi^2m_s^4}{16c_H^2\mu^4}\right)^{1/2}\nonumber\\
&&\times\left(\frac{\mu}{2.6\mbox{ TeV}}\right)^2
\left(\frac{10^{15}\mbox{ GeV}}{f_a}\right)
\left(\frac{450\mbox{ GeV}}{m_s}\right)^{1/2},
\end{eqnarray}
and radiation-saxion equality temperature,
\begin{eqnarray}
T_e^s=\frac43m_sY_s^{\rm CO}=2.5\mbox{ GeV}
\left(\frac{{\rm min}[T_R,T_s]}{10^7\mbox{ GeV}}\right)
\left(\frac{f_a}{10^{15}\mbox{ GeV}}\right)^2.
\end{eqnarray}
Notice that $r$ can be as large as ${\cal O}(1000)$ for $f_a\sim 10^{16}$ GeV as shown in Fig.~\ref{fig:DNeff_SOA}.
 Therefore, the neutralino density can be ${\cal O}(1/1000)$ times smaller than the standard density as shown
in Fig.~\ref{fig:neut_rean_SOA} and so the overproduction constraint can be avoided.
The saxion decay temperature is around 10 MeV even for $f_a\sim 10^{16}$ GeV leaving unchanged
the standard BBN prediction.

\section{Conclusion}

The supersymmetric DFSZ axion model is highly motivated in that it provides 
1.\ the SUSY solution to the gauge hierarchy problem, 
2.\ the Peccei-Quinn-Weinberg-Wilczek solution to the strong $CP$ problem and  
3.\ the Kim-Nilles solution to the SUSY $\mu$ problem. 
We examined production rates for mixed axion/neutralino dark matter within the SUSY DFSZ model for a standard underabundance model (SUA) and a standard overabundance model (SOA).
Much of the cosmology depends on the axino and saxion decay modes which are very different than those expected from the SUSY KSVZ model. In SUSY DFSZ, the direct coupling of axinos and saxions 
to the Higgs supermultiplets allows for rapid decays into various Higgs-Higgs, Higgs-Higgsino and diboson final states which do not occur in the SUSY KSVZ model.

For the SUA case (which has low $\mu\sim 150$ GeV as required by naturalness), the dark matter scenarios broke up into three main cases. For the lower range of $f_a\alt 10^{12}$ GeV, axinos and saxions can be thermally produced, but decay before the neutralino freeze-out, so that the standard relic neutralino abundance holds true. In this case of underabundant neutralinos, the remaining dark matter is composed of axions. For SUA, Higgsino-like WIMPs comprise $\sim 10\%$ of the CDM while axions comprise $\sim 90\%$. Dark radiation from $s\to aa$ decay is scant.
For $f_a\sim 10^{12}-10^{14}$ GeV, saxions and axinos do not dominate the universe, but do decay after the neutralino freeze-out, augmenting the standard abundance.
The remaining axion abundance can always be adjusted using the initial misalignment $\theta_i$ so that the total mixed axion/neutralino abundance saturates $\Omega_{a\tz_1}h^2\sim 0.12$ as long as neutralinos are not overproduced. For $f_a\agt 10^{14}$ GeV, oscillation-produced
saxions may dominate the universe and can overproduce both neutralino dark matter and dark radiation. However, in cases where $m_s$ is light enough to forbid saxion decays to 
SUSY particles, and where $\xi$ is small enough to suppress dark radiation, 
saxion decay to SM particles can lead to large entropy dilution of neutralinos and axions
so that the measured CDM abundance can be obtained.

\begin{figure}
\begin{center}
\includegraphics[height=6.9cm]{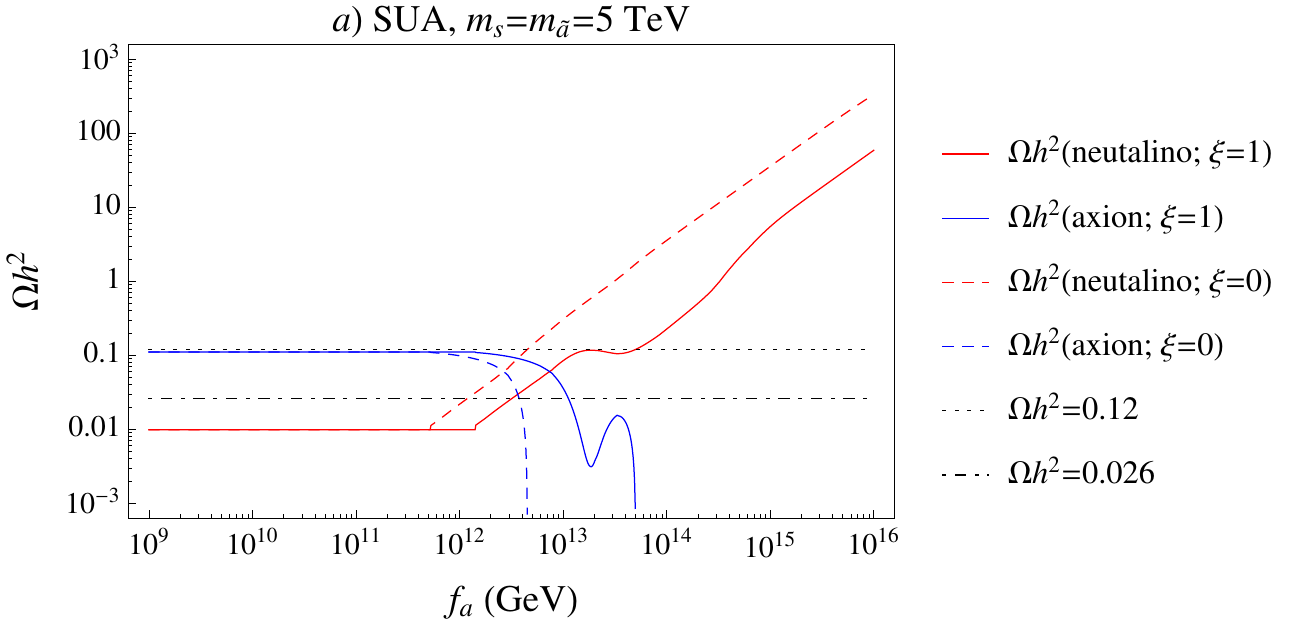}~~~~~\\
\includegraphics[height=6.9cm]{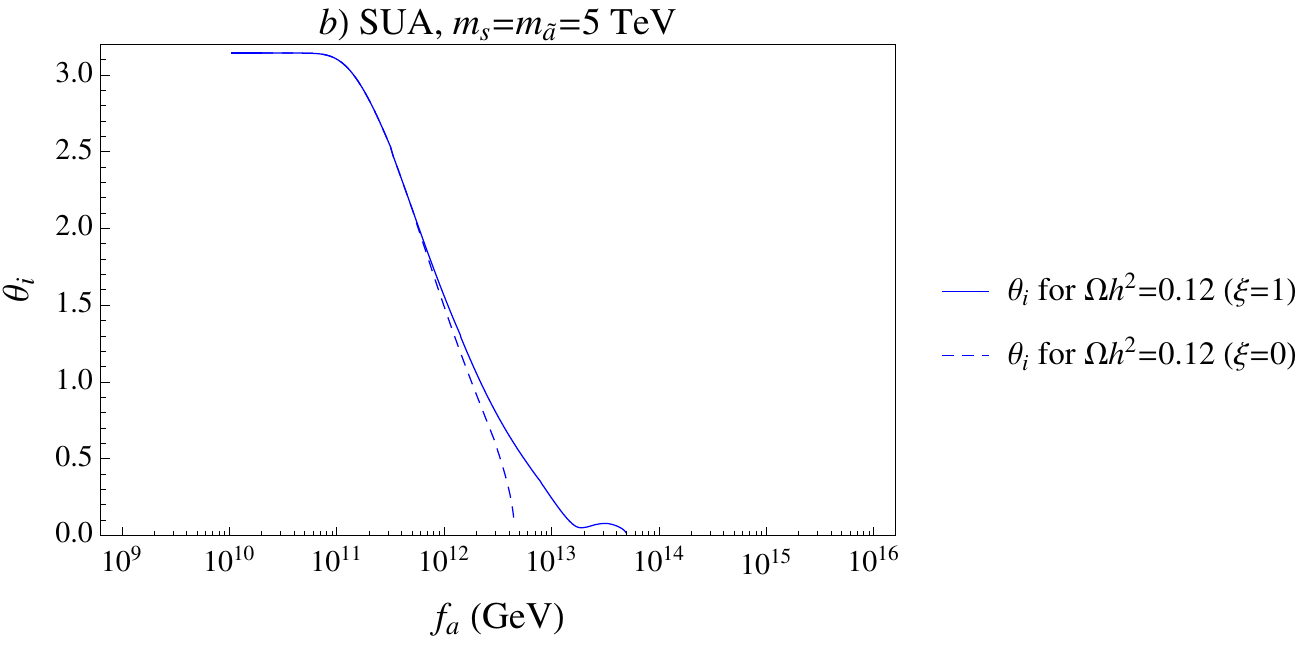}
\caption{Plot of {\it a}) $\Omega_{\tz_1}h^2$ and $\Omega_ah^2$ vs. $f_a$ for the SUA benchmark
with $m_s=m_{\ta}=5$ TeV for $\xi= 0$ (dashed) and $\xi=1$ (solid).
In {\it b}), we show the required axion misalignment angle $\theta_i$ required to
saturate the mixed axion/neutralino abundance to match the measured value.
\label{fig:SUA_5tev}}
\end{center}
\end{figure}

\begin{figure}
\begin{center}
\includegraphics[height=6.9cm]{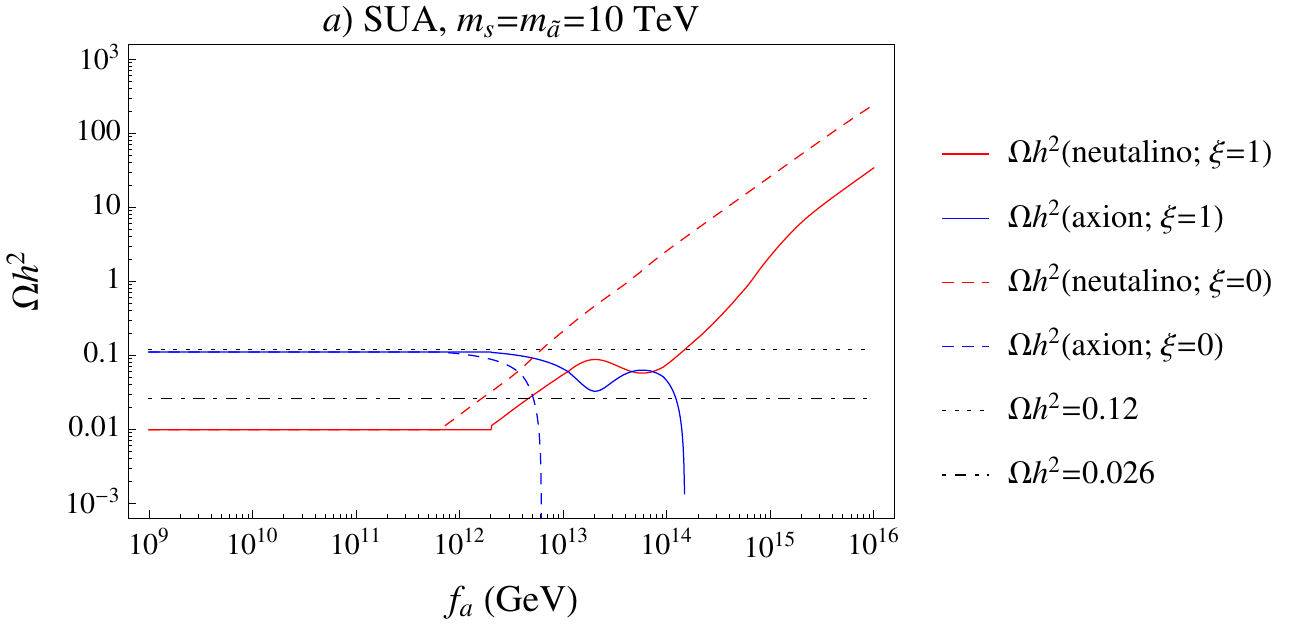}~~~~~\\
\includegraphics[height=6.9cm]{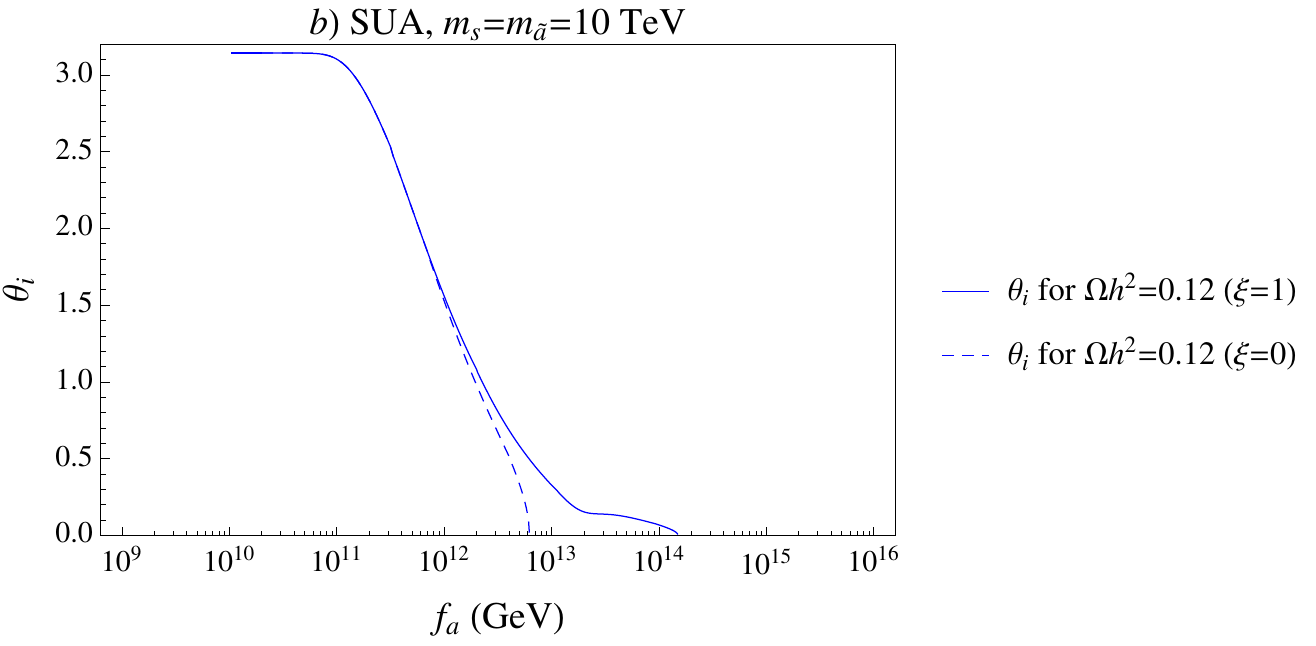}
\caption{Plot of {\it a}) $\Omega_{\tz_1}h^2$ and $\Omega_ah^2$ vs. $f_a$ for the SUA benchmark
with $m_s=m_{\ta}=10$ TeV for $\xi= 0$ (dashed) and $\xi=1$ (solid).
In {\it b}), we show the required axion misalignment angle $\theta_i$ required to
saturate the mixed axion/neutralino abundance to match the measured value.
\label{fig:SUA_10tev}}
\end{center}
\end{figure}

\begin{figure}
\begin{center}
\includegraphics[height=6.9cm]{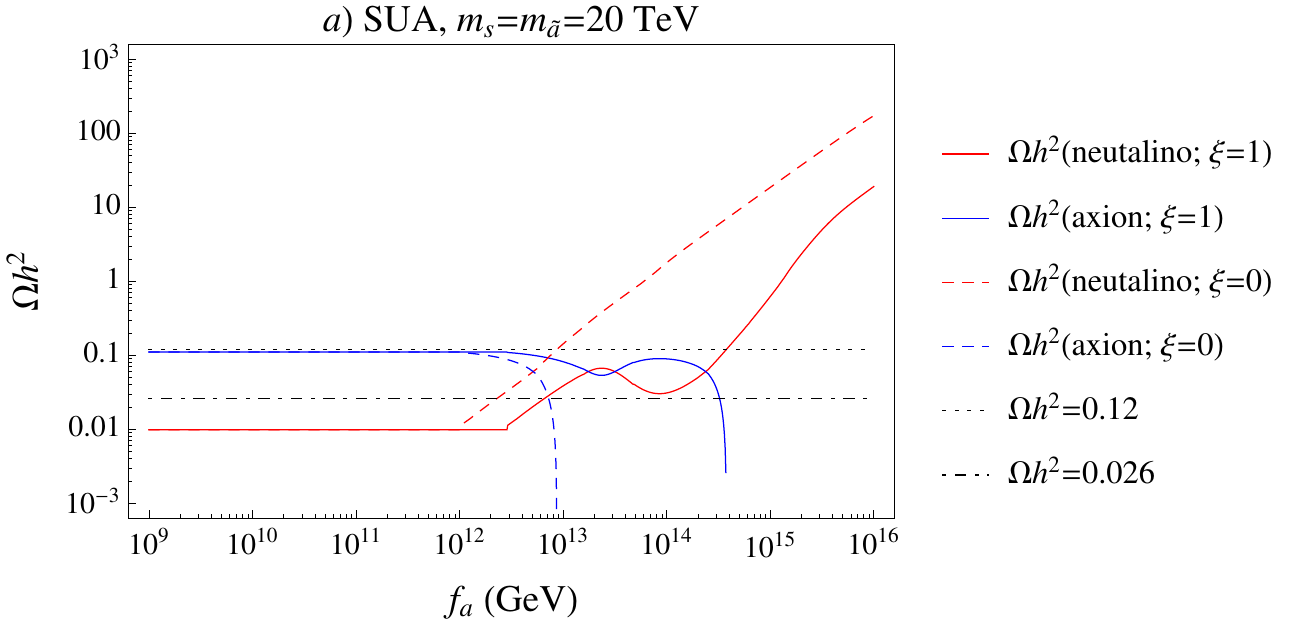}~~~~~\\
\includegraphics[height=6.9cm]{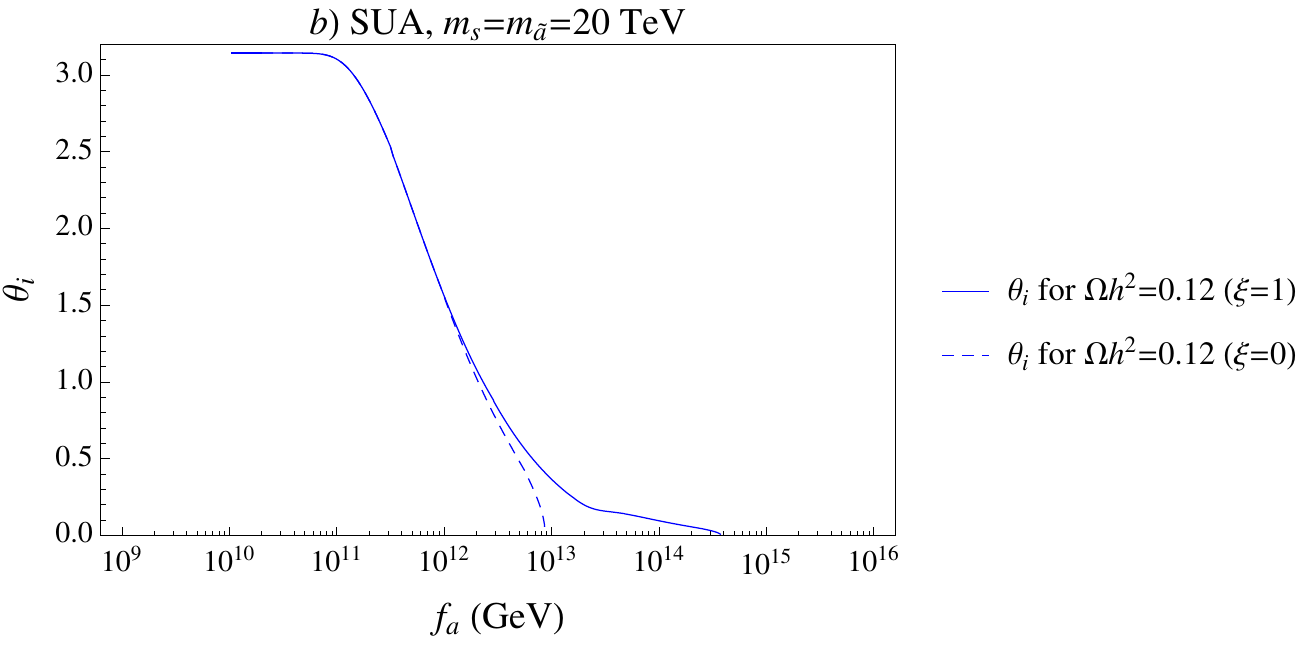}
\caption{Plot of {\it a}) $\Omega_{\tz_1}h^2$ and $\Omega_ah^2$ vs. $f_a$ for the SUA benchmark
with $m_s=m_{\ta}=20$ TeV for $\xi= 0$ (dashed) and $\xi=1$ (solid).
In {\it b}), we show the required axion misalignment angle $\theta_i$ required to
saturate the mixed axion/neutralino abundance to match the measured value.
\label{fig:SUA_20tev}}
\end{center}
\end{figure}

Our summary plots for the SUA case are given in Figs.~\ref{fig:SUA_5tev}, \ref{fig:SUA_10tev} and \ref{fig:SUA_20tev} where the panels {\it a}) show the values of $\Omega_{\tz_1}h^2$ and 
$\Omega_a h^2$ versus $f_a$ for $m_s=m_{\ta}=5$ TeV, 10 TeV and 20 TeV, respectively.
Dashed curves are for $\xi =0$ while solid curves are for $\xi =1$.
In panels {\it b}), we show the required value of the
axion misalignment angle $\theta_i$ with which the total neutralino plus axion abundance saturates
the measured value. From Fig.~\ref{fig:SUA_5tev}{\it a}), we see-- over the large range of
$f_a\sim 10^9-10^{12}$ GeV--  
that Higgsino-like WIMPs comprise just $\sim 10\%$ of the total dark matter abundance, 
while the remaining 90\% is comprised of axions.
This region of mainly axion CDM from natural SUSY models has been emphasized in Ref.~\cite{prl}.

For SUA, the spin-independent (SI) neutralino-proton scattering cross section is $\sigma^{\rm SI}(\tz_1p)\simeq 1.7\times 10^{-8}$ pb as shown in Table~\ref{tab:bm}, whilst the limit from 225 live days of Xe-100 data taking~\cite{xenon} is $\sigma^{\rm SI}(\tz_1p)\alt 4\times 10^{-9}$ pb for a 135 GeV WIMP. This apparent conflict is easily reconciled within the
SUA benchmark  as the relic Higgsino-like WIMPs comprise only a fraction of the local relic density, and so the Xe-100 limits have to be rescaled downward by a factor $\Omega_{\tz_1}h^2/0.12$. 
For SUA with $f_a\alt 10^{12}$ GeV, the rescaling factor is $\sim 0.1$. 
The rescaled SI Higgsino-like WIMP detection rates compared against limits have been shown in
Ref.~\cite{bbm} for a variety of radiatively-driven natural SUSY models.  In Figs.~\ref{fig:SUA_5tev}, \ref{fig:SUA_10tev} and \ref{fig:SUA_20tev}, also shown are
the lines for $\Omega_{\widetilde{Z}_1}h^2 =0.026$ below which the Xe-100 bound is evaded.

For the $\xi=0$ case of Fig.~\ref{fig:SUA_5tev}{\it a}),
the $\Omega_{\tz_1}h^2$ curve  rises steadily with large $f_a\agt 10^{12}$ GeV due to increasing production of
saxions from coherent oscillations and their dominant decays to SUSY particles. This leads to subsequent neutralino
re-annihilation at decreasing temperatures $T_D^s$. For $\xi =1$, the dominant saxion decay mode is $s\to aa$,
and decay-produced neutralinos come mainly from thermal axino production which decreases as
$f_a$ increases. One sees that $\Omega_{\tz_1}h^2$ turns over and briefly reaches $\Omega_{\tz_1}h^2\simeq 0.1$
at $f_a\sim 3\times 10^{13}$ GeV before beginning again a rise due to increasing non-thermal saxion
production. It is important to note that for  $\xi \sim 1$ and $f_a\agt 10^{14}$ GeV,  too much dark radiation
is produced ($\Delta N_{eff}>1.6$, see Fig.~\ref{fig:DNeff_SUA_ms=5000}) and thus very large $f_a$ is excluded by
overproduction of both dark radiation and WIMPs.

In all cases shown,  axino and saxion decay widths become suppressed
and they decay after neutralino freeze-out leading to an augmented neutralino abundance 
as $f_a$ increases beyond $10^{12}$ GeV. In this region, WIMP dark matter becomes overproduced 
and the model becomes excluded. The excluded region occurs at higher $f_a$ for $\xi=1$ models since these cases allow for $s\to aa$ decay which tends to be the dominant saxion decay mode when it is fully allowed; in such cases, the $f_a$ value at which $T_D$ drops
below $T_{fr}$ increases.

The axion misalignment angle shown in panels {\it b}) is required to be nearly 
$\theta_i\sim \pi$ for low $f_a\alt 10^{11}$ GeV
but more natural values of $\theta_i$ occur for $f_a\sim 10^{11}-10^{13}$ GeV.
\footnote{Here we discuss with $f_a$ not with $f_a/N_{\rm DW}$. Since the axion CO density is determined by $f_a/N_{\rm DW}$, so the natural range of $f_a$ is about an order of magnitude larger than that in the KSVZ model. }
As $m_{\ta}=m_s$ is increased to 10 (20) TeV, the upper bound on $f_a$ moves to $6\times 10^{12}$
($8\times 10^{12}$) GeV for $\xi=0$ as shown in Figs.~\ref{fig:SUA_10tev} and \ref{fig:SUA_20tev}.
In the case of $\xi=1$, there is a window of $\Omega_{\widetilde{Z}_1}h^2<0.12$ in the region of
$10^{13}$ GeV$\lesssim f_a\lesssim 10^{14}$ GeV for $m_{\ta}=m_s\sim10-20$ TeV.
But, it is still above the Xe-100 limit, {\it i.e.} $\Omega_{\widetilde{Z}_1}h^2>0.026$, and thus
this region is excluded within the SUA benchmark scenario.

\begin{figure}
\begin{center}
\includegraphics[height=6.9cm]{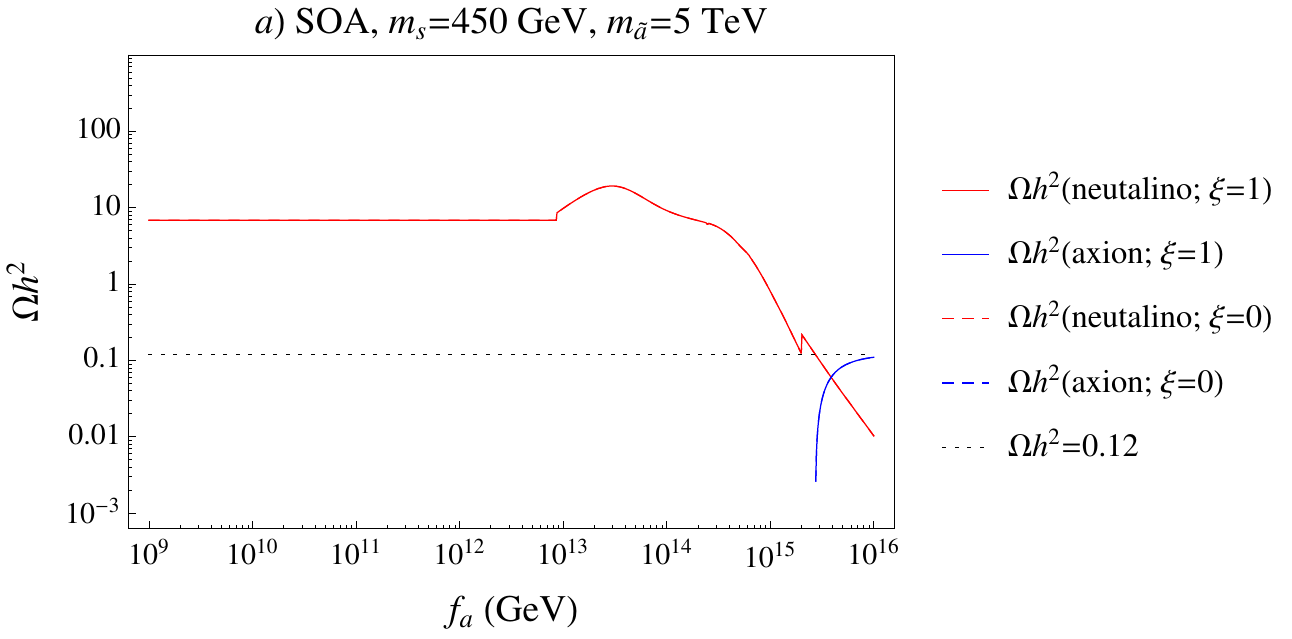}~~~~\\
\includegraphics[height=6.9cm]{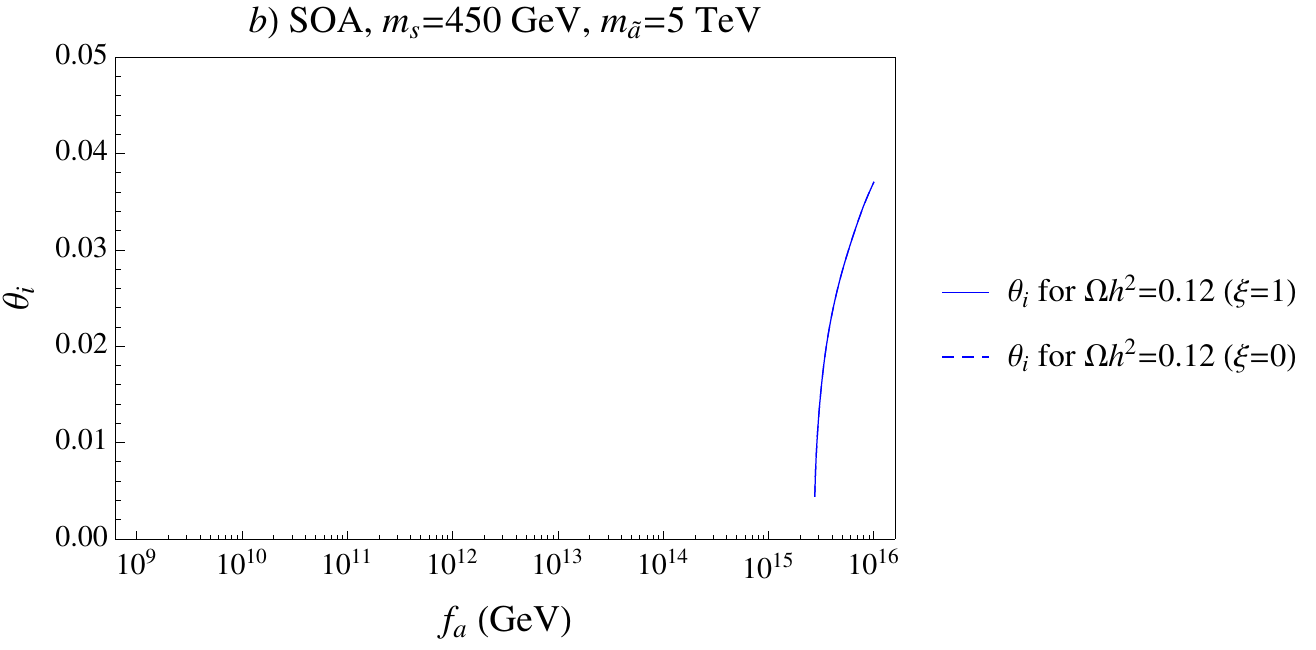}
\caption{Plot of {\it a}) $\Omega_{\tz_1}h^2$ and $\Omega_ah^2$ vs. $f_a$ for the SOA benchmark
with $m_s=m_{\ta}=0.5$ TeV for $\xi= 0$ (dashed) and $\xi=1$ (solid).
In {\it b}), we show the required axion misalignment angle $\theta_i$ required to
saturate the mixed axion/neutralino abundance to match the measured value.
Plots of $\xi=1,0$ cases are overlapped since $BR(s\to aa)$ is very tiny for small $m_s$ region.
\label{fig:SOA}}
\end{center}
\end{figure}

For the SOA case with low $f_a\alt 10^{13}$ GeV, the standard neutralino abundance remains
unchanged, and thus the model is excluded by the overabundance. 
As $f_a$ is raised to $10^{14}$ GeV,  saxions and axinos decay after the
freeze-out, augmenting the neutralino overabundance even further. 
For $f_a\agt 10^{14}$ GeV, as in the SUA case, the universe is dominated by 
oscillation-produced saxions leading to injection of even more neutralino dark matter 
as well as dark radiation. Thus, a large range of $f_a$ is excluded in SOA 
by overproduction of dark matter, and also possibly by overproduction of dark radiation.
An exception occurs for small $m_s$ and low $\xi$ where saxion decays to SUSY particles and axions are suppressed, and large entropy injection can bring the combined neutralino and axion abundance into accord with measured values. This is the case for $f_a\sim 3\times 10^{15}$ GeV  shown in Fig.~\ref{fig:SOA} where $m_s=450$ GeV, $m_{\ta}=5$ TeV is selected to close most of the lucrative saxion decay modes to SUSY particles. The axion abundance is suppressed appropriately by both entropy dilution and a small value of $\theta_i$ as shown in panel {\it b}).

{\it Summary:} We have considered $R$-parity conserving SUSY models with a standard under- and 
over-abundance of dark matter which invoke the PQWW solution to the strong CP problem via the SUSY DFSZ model,
wherein Higgs superfields carry PQ charge, and which also provides a solution to the SUSY $\mu$ problem. 
For standard underabundant models, over a large range of PQ scale $f_a\sim 10^9-10^{12}$ GeV,
saxions and axinos typically decay before neutralino freeze-out so that the WIMP portion of dark matter 
is expected to lie at its standard predicted value from thermal freeze-out, 
while axions would comprise the remainder. 
The relic neutralinos stand a good chance to be detectable at next generation WIMP direct detection 
experiments even with a depleted local abundance. 
Prospects for WIMP indirect detection should be more limited since expected rates go as the 
depleted abundance squared~\cite{bbm}. Prospects for microwave cavity detection of axions are 
good for the range of $f_a$ where mainly axion dark matter is expected; 
in such cases, axions should be accessible to experimental searches~\cite{axsearch}. 
For standard overabundant models, on the other hand, overabundant neutralino dark matter density 
can be appropriately depleted by a large entropy production from the oscillation-produced saxion decay. 
In this case, prospects for detecting both WIMP and axion dark matter are not promising.

\acknowledgments

We thank A. Lessa for ongoing collaboration on these topics.

\appendix
\section{Appendix: partial decay widths of saxion and axino}

We show the exact partial decay widths of saxion and axino at the tree-level.
All the conventions are as in Ref.~\cite{BaerTata}.
\begin{itemize}

\item $s\to \phi_i\phi_j$
\begin{equation}
\Gamma(s\to\phi_i\phi_j)=\frac{\lambda_{s\phi_i\phi_j}^2}{16\pi m_s}
\lambda^{1/2}\left(1,\frac{m_{\phi_i}^2}{m_s^2},\frac{m_{\phi_j}^2}{m_s^2}\right)
\left(1-\frac12\delta_{ij}\right),
\label{dec_sphiphi}
\end{equation}
where $\phi_i=h,H,A,H^+,H^-$.
Trilinear couplings are given by
\begin{eqnarray}
\lambda_{shh}&=&\left(\frac{\sqrt{2}c_H\mu^2}{v_{PQ}}\right)
\left\{1-\frac14\left(\frac{m_A^2}{\mu^2}\right)\sin2\beta\sin2\alpha\right\}\nonumber\\
&&+\left(\frac{M_Z^2}{\sqrt{2}v}\right)\cos2\alpha\sin(\beta-\alpha)\left[3\epsilon_h-\epsilon_H\left\{2\tan2\alpha+\cot(\beta-\alpha)\right\}\right],
\label{int_shh}\\
\lambda_{sHH}&=&\left(\frac{\sqrt{2}c_H\mu^2}{v_{PQ}}\right)
\left\{1+\frac14\left(\frac{m_A^2}{\mu^2}\right)\sin2\beta\sin2\alpha\right\}\nonumber\\
&&+\left(\frac{M_Z^2}{\sqrt{2}v}\right)\cos2\alpha\sin(\beta-\alpha)\left[3\epsilon_H\cot(\beta-\alpha)
\right.\nonumber\\
&&
\left.+\epsilon_h\left\{2\tan2\alpha\cot(\beta-\alpha)-1\right\}\right],
\label{int_sHH}\\
\lambda_{shH}&=&-\left(\frac{c_Hm_A^2}{2\sqrt{2}v_{PQ}}\right)\sin2\beta\cos2\alpha\nonumber\\
&&+\left(\frac{M_Z^2}{\sqrt{2}v}\right)\cos2\alpha\sin(\beta-\alpha)
\left[
-\epsilon_h\left\{2\tan2\alpha+\cot(\beta-\alpha)\right\}\right.\nonumber\\
&&\left.+\epsilon_H\left\{2\tan2\alpha\cot(\beta-\alpha)-1\right\}
\right],
\label{int_shH}\\
\lambda_{sAA}&=&\left(\frac{\sqrt{2}c_H\mu^2}{v_{PQ}}\right)
\left\{1+\frac14\left(\frac{m_A^2}{\mu^2}\right)\sin^22\beta\right\}\nonumber\\
&&+\left(\frac{M_Z^2}{\sqrt{2}v}\right)\cos2\beta\sin(\beta-\alpha)
\left\{
\epsilon_h-\epsilon_H\cot(\beta-\alpha)
\right\},
\label{int_sAA}\\
\lambda_{sH^+H^-}&=&\left(\frac{\sqrt{2}c_H\mu^2}{v_{PQ}}\right)
\left\{1+\frac14\left(\frac{m_A^2}{\mu^2}\right)\sin^22\beta\right\}\nonumber\\
&&+\left(\frac{M_Z^2}{\sqrt{2}v}\right)
\left[
\cos2\beta\sin(\beta-\alpha)\left\{\epsilon_h-\epsilon_H\cot(\beta-\alpha)\right\}\right.\nonumber\\
&&\left.+2\cos^2\theta_W\sin(\beta+\alpha)\left\{\epsilon_h+\epsilon_H\cot(\beta+\alpha)\right\}
\right].
\label{int_sH+H-}
\end{eqnarray}

\item $s\to VV$

\begin{eqnarray}
\Gamma(s\to VV)&=&\frac{g_V^2g_{sVV}^2}{16\pi}m_s\left\{3\frac{M_V^2}{m_s^2}+\frac{m_s^2}{4M_V^2}
\left(1-\frac{4M_V^2}{m_s^2}\right)\right\}
\nonumber\\
&&\times\left(1-\frac{4M_V^2}{m_s^2}\right)^{1/2}\left(1-\frac12\delta_{VZ}\right)
\end{eqnarray}
with
\begin{equation}
g_{sVV}=\epsilon_hg_{hVV}+\epsilon_Hg_{HVV}=\epsilon_h\sin(\beta+\alpha)+\epsilon_H\cos(\beta+\alpha),
\end{equation}
where $\delta_{VZ}=1,0$ for $V=Z,W$ and $g_W=g$ and $g_Z=g/\cos\theta_W$.
$\epsilon_h$ and $\epsilon_H$ are the saxion-Higgs mixing components, which are given by
\begin{eqnarray}
\epsilon_h&=&
\left(\frac{1}{m_h^2-m_s^2}\right)\left(\frac{v\sin2\beta}{2v_{PQ}}\right)
\left\{
m_A^2\cos(\beta-\alpha)-\frac{4\mu^2\sin(\alpha+\beta)}{\sin2\beta}
\right\},
\label{sh_mixing1}\\
\epsilon_H&=&
\left(\frac{1}{m_H^2-m_s^2}\right)\left(\frac{v\sin2\beta}{2v_{PQ}}\right)
\left\{
m_A^2\sin(\beta-\alpha)-\frac{4\mu^2\cos(\alpha+\beta)}{\sin2\beta}
\right\}.\label{sh_mixing2}
\end{eqnarray}

\item $s\to f\bar{f}$
\begin{equation}
\Gamma(s\to f\bar{f})=\frac{N_c}{16\pi}\frac{m_f^2}{v^2}g_{sff}^2m_s\left(1-\frac{4m_f^2}{m_s^2}\right)^{3/2}
\end{equation}
with
\begin{eqnarray}
g_{sff}&=&\epsilon_hg_{hff}+\epsilon_Hg_{Hff}\nonumber\\
&=&
\left\{
\begin{array}{ll}
\frac{1}{\sin\beta}\left(-\epsilon_h\cos\alpha+\epsilon_H\sin\alpha\right), &\mbox{for up-type fermions},\\
\frac{1}{\cos\beta}\left(-\epsilon_h\sin\alpha-\epsilon_H\cos\alpha\right), & \mbox{for down-type fermions}.
\end{array}
\right.
\end{eqnarray}

\item $s\to\widetilde{Z}_i\widetilde{Z}_j~/~\widetilde{W}_i\widetilde{W}_j$

\begin{eqnarray}
\Gamma(s\to\widetilde{W}_i^+\widetilde{W}_i^-)
&=&\frac{g^2}{4\pi}\left|\Sigma_i^s\right|^2m_s
\left(1-\frac{4m_{\widetilde{W}_i}^2}{m_s^2}\right)^{3/2},\\
\label{dec_scharii}
\Gamma(s\to\widetilde{W}_1^+\widetilde{W}_2^-)
&=&
\Gamma(s\to\widetilde{W}_1^-\widetilde{W}_2^+)\nonumber\\
&=&
\frac{g^2}{16\pi}m_s
\lambda^{1/2}\left(
1,\frac{m_{\widetilde{W}_1}^2}{m_s^2},\frac{m_{\widetilde{W}_2}^2}{m_s^2}
\right)\\
&&\times\left[
\left|\Sigma^s\right|^2\left\{1-
\left(\frac{m_{\widetilde{W}_2}+m_{\widetilde{W}_1}}{m_s}\right)^2
\right\}\right.\nonumber\\
&&\left.+\left|\Pi^s\right|^2\left\{1-
\left(\frac{m_{\widetilde{W}_2}-m_{\widetilde{W}_1}}{m_s}\right)^2
\right\}\right].\label{dec_schar12}\\
\Gamma(s\to\widetilde{Z}_i\widetilde{Z}_j)&=
&\frac{1}{8\pi}m_s\left(\Xi^s_{ij}+\Xi^s_{ji}\right)^2
\left[
1-\left\{\frac{m_{\widetilde{Z}_i}+(-1)^{\theta_i+\theta_j}m_{\widetilde{Z}_j}}{m_s}\right\}^2
\right]\nonumber\\
&&\times\lambda^{1/2}\left(
1,\frac{m_{\widetilde{Z}_i}^2}{m_s^2},\frac{m_{\widetilde{Z}_j}^2}{m_s^2}\right)
\left(1-\frac12\delta_{ij}\right),\label{dec_sneut}
\end{eqnarray}
where $\theta_i$ is 1(0) if $i$-th eigenvalue of neutralino mass matrix is negative (positive) and $m_{\widetilde{Z}_i}$ is always positive.
$\theta_{\widetilde{W}_i}$ is 1(0) if $i$-th eigenvalue of chargino mass matrix is negative (positive) and $m_{\widetilde{W}_i}$ is always positive.
The neutralino couplings are given by
\begin{eqnarray}
\Sigma^s_1&=&\epsilon_hS_1^h+\epsilon_HS_1^H-\frac{c_H\mu}{4gv_{PQ}}S^s_1,\\
\Sigma^s_2&=&\epsilon_hS_2^h+\epsilon_HS_2^H-\frac{c_H\mu}{4gv_{PQ}}S^s_2,\\
\Sigma^s&=&\epsilon_hS^h+\epsilon_HS^H-\frac{c_H\mu}{2gv_{PQ}}S^s,\\
\Pi^s&=&\epsilon_hP^h+\epsilon_HP^H-\frac{c_H\mu}{2gv_{PQ}}P^s,\\
\Xi^s_{ij}&=&\epsilon_hX_{ij}^h+\epsilon_HX_{ij}^H-\frac{c_H\mu}{2\sqrt{2}v_{PQ}}X^s_{ij},
\end{eqnarray}
where
\begin{eqnarray}
X^s_{ij}&=&(-1)^{\theta_i+\theta_j}v_1^{(i)}v_2^{(j)},\\
S_1^s&=&(-1)^{\theta_{\widetilde{W}_1}}\cos\gamma_L\cos\gamma_R,
\\
S_2^s&=&-(-1)^{\theta_{\widetilde{W}_2}}\theta_x\theta_y\sin\gamma_L\sin\gamma_R,
\\
S^s&=&\frac12\left\{
(-1)^{\theta_{\widetilde{W}_1}}\theta_y\cos\gamma_L\sin\gamma_R
-(-1)^{\theta_{\widetilde{W}_2}}\theta_x\sin\gamma_L\cos\gamma_R
\right\},
\\
P^s&=&\frac12\left\{
(-1)^{\theta_{\widetilde{W}_1}}\theta_y\cos\gamma_L\sin\gamma_R
+(-1)^{\theta_{\widetilde{W}_2}}\theta_x\sin\gamma_L\cos\gamma_R
\right\},
\end{eqnarray}
and
\begin{eqnarray}
S_1^h&=&\frac12(-1)^{\theta_{\widetilde{W}_1}}
\left[\sin\alpha\sin\gamma_R\cos\gamma_L+\cos\alpha\sin\gamma_L\cos\gamma_R\right],\\
S_2^h&=&\frac12(-1)^{\theta_{\widetilde{W}_2}+1}\theta_x\theta_y
\left[\sin\alpha\cos\gamma_R\sin\gamma_L+\cos\alpha\cos\gamma_L\sin\gamma_R\right],\\
S^h&=&\frac12\left[
-(-1)^{\theta_{\widetilde{W}_1}}\theta_x\sin\gamma_R\sin\gamma_L\sin\alpha
+(-1)^{\theta_{\widetilde{W}_1}}\theta_x\cos\gamma_R\cos\gamma_L\cos\alpha
\right.\nonumber\\
&&\left.-(-1)^{\theta_{\widetilde{W}_2}}\theta_y\sin\gamma_R\sin\gamma_L\cos\alpha
+(-1)^{\theta_{\widetilde{W}_2}}\theta_y\cos\gamma_R\cos\gamma_L\sin\alpha
\right],\\
P^h&=&\frac12\left[
+(-1)^{\theta_{\widetilde{W}_1}}\theta_x\sin\gamma_R\sin\gamma_L\sin\alpha
-(-1)^{\theta_{\widetilde{W}_1}}\theta_x\cos\gamma_R\cos\gamma_L\cos\alpha
\right.\nonumber\\
&&\left.-(-1)^{\theta_{\widetilde{W}_2}}\theta_y\sin\gamma_R\sin\gamma_L\cos\alpha
+(-1)^{\theta_{\widetilde{W}_2}}\theta_y\cos\gamma_R\cos\gamma_L\sin\alpha
\right],\\
X_{ij}^h&=&-\frac12(-1)^{\theta_i+\theta_j}
\left(v_2^{(i)}\sin\alpha-v_1^{(i)}\cos\alpha\right)
\left(gv_3^{(j)}-g'v_4^{(j)}\right).
\label{eq:Xhij}
\end{eqnarray}
The couplings of the heavy scalar $H$ can be obtained from those of $h$ by replacing $\cos\alpha\to-\sin\alpha$ and $\sin\alpha\to\cos\alpha$.
$v_i^{(j)}$ is the neutralino mixing component and $\gamma_{L,R}$ is chargino mixing angle.
They are defined in Ref.~\cite{BaerTata}.

\item $s\to\tilde{f}_i\tilde{f}_j$

\begin{equation}
\Gamma(s\to\tilde{f}_i\tilde{f}_j)=\frac{\left|\epsilon_h{\cal A}^h_{\tilde{f}_i\tilde{f}_j}+\epsilon_H{\cal A}^H_{\tilde{f}_i\tilde{f}_j}\right|^2}{16\pi m_s}N_c(f)
\lambda^{1/2}\left(1,\frac{m_{\tilde{f}_i}^2}{m_s^2},\frac{m_{\tilde{f}_j}^2}{m_s^2}\right),
\end{equation}
where
\begin{eqnarray}
{\cal A}_{\tilde{f}_1\tilde{f}_1}^{h,H}&=&
{\cal A}_{\tilde{f}_L\tilde{f}_L}^{h,H}\cos^2\theta_f
+{\cal A}_{\tilde{f}_R\tilde{f}_R}^{h,H}\sin^2\theta_f
-2{\cal A}_{\tilde{f}_L\tilde{f}_R}^{h,H}\cos\theta_f\sin\theta_f,\\
{\cal A}_{\tilde{f}_2\tilde{f}_2}^{h,H}&=&
{\cal A}_{\tilde{f}_L\tilde{f}_L}^{h,H}\sin^2\theta_f
+{\cal A}_{\tilde{f}_R\tilde{f}_R}^{h,H}\cos^2\theta_f
+2{\cal A}_{\tilde{f}_L\tilde{f}_R}^{h,H}\cos\theta_f\sin\theta_f,\\
{\cal A}_{\tilde{f}_1\tilde{f}_2}^{h,H}&=&
{\cal A}_{\tilde{f}_L\tilde{f}_L}^{h,H}\cos\theta_f\sin\theta_f
-{\cal A}_{\tilde{f}_R\tilde{f}_R}^{h,H}\cos\theta_f\sin\theta_f
+2{\cal A}_{\tilde{f}_L\tilde{f}_R}^{h,H}\cos2\theta_f,\\
{\cal A}_{\tilde{f}_2\tilde{f}_1}^{h,H}&=&{\cal A}_{\tilde{f}_1\tilde{f}_2}^{h,H}
\end{eqnarray}
with
\begin{eqnarray}
{\cal A}_{\tilde{u}_L\tilde{u}_L}^h&=&
g\left[
M_W\left(\frac12-\frac16\tan^2\theta_W\right)\sin(\beta-\alpha)-\frac{m_u^2\cos\alpha}{M_W\sin\beta}
\right],\\
{\cal A}_{\tilde{u}_R\tilde{u}_R}^h&=&
g\left[
\frac23M_W\tan^2\theta_W\sin(\beta-\alpha)-\frac{m_u^2\cos\alpha}{M_W\sin\beta}
\right],\\
{\cal A}_{\tilde{u}_L\tilde{u}_R}^h&=&
\frac{gm_u}{2M_W\sin\beta}\left(-\mu\sin\alpha+A_u\cos\alpha\right),
\end{eqnarray}
\begin{eqnarray}
{\cal A}_{\tilde{u}_L\tilde{u}_L}^H&=&
g\left[
-M_W\left(\frac12-\frac16\tan^2\theta_W\right)\cos(\beta-\alpha)+\frac{m_u^2\sin\alpha}{M_W\sin\beta}
\right],\\
{\cal A}_{\tilde{u}_R\tilde{u}_R}^H&=&
g\left[
-\frac23M_W\tan^2\theta_W\cos(\beta-\alpha)+\frac{m_u^2\sin\alpha}{M_W\sin\beta}
\right],\\
{\cal A}_{\tilde{u}_L\tilde{u}_R}^H&=&
\frac{gm_u}{2M_W\sin\beta}\left(-\mu\cos\alpha-A_u\sin\alpha\right),
\end{eqnarray}
\begin{eqnarray}
{\cal A}_{\tilde{d}_L\tilde{d}_L}^h&=&
g\left[
M_W\left(-\frac12-\frac16\tan^2\theta_W\right)\sin(\beta-\alpha)-\frac{m_d^2\sin\alpha}{M_W\cos\beta}
\right],\\
{\cal A}_{\tilde{d}_R\tilde{d}_R}^h&=&
g\left[
-\frac13M_W\tan^2\theta_W\sin(\beta-\alpha)-\frac{m_d^2\sin\alpha}{M_W\cos\beta}
\right],\\
{\cal A}_{\tilde{d}_L\tilde{d}_R}^h&=&
\frac{gm_d}{2M_W\cos\beta}\left(-\mu\cos\alpha+A_d\sin\alpha\right),
\end{eqnarray}
\begin{eqnarray}
{\cal A}_{\tilde{d}_L\tilde{d}_L}^H&=&
g\left[
M_W\left(\frac12+\frac16\tan^2\theta_W\right)\cos(\beta-\alpha)-\frac{m_d^2\cos\alpha}{M_W\cos\beta}
\right],\\
{\cal A}_{\tilde{d}_R\tilde{d}_R}^H&=&
g\left[
\frac13M_W\tan^2\theta_W\cos(\beta-\alpha)-\frac{m_d^2\cos\alpha}{M_W\cos\beta}
\right],\\
{\cal A}_{\tilde{d}_L\tilde{d}_R}^H&=&
\frac{gm_d}{2M_W\cos\beta}\left(\mu\sin\alpha+A_d\cos\alpha\right).
\end{eqnarray}
$\theta_f$ is sfermion mixing angle, which is defined by Ref.~\cite{BaerTata}.

\item $\tilde{a}\to\widetilde{Z}_i\phi$

\begin{eqnarray}
\Gamma(\tilde{a}\to\widetilde{Z}_i\phi)&=&\frac{1}{16\pi}
\left(\Lambda_{\tilde{a}\widetilde{Z}\phi}^i\right)^2
m_{\tilde{a}}
\lambda^{1/2}\left(1,\frac{m_{\widetilde{Z}_i}^2}{m_{\tilde{a}}^2},\frac{m_{\phi}^2}{m_{\tilde{a}}^2}\right)\nonumber\\
&&\times\left[
\left(1+\frac{m_{\widetilde{Z}_i}^2}{m_{\tilde{a}}^2}-\frac{m_{\phi}^2}{m_{\tilde{a}}^2}\right)
+2(-1)^{\theta_i+\theta_{\tilde{a}}}\left(1-2\delta_{A\phi}\right)\frac{m_{\widetilde{Z}_i}}{m_{\tilde{a}}}
\right],
\label{dec_aZphi}
\end{eqnarray}
for $\phi=h,H,A$.
The couplings are given by
\begin{eqnarray}
\Lambda_{\tilde{a}\tilde{Z}h}^i&=&X_{i0}^h+X_{0i}^h-\frac{c_H\mu}{\sqrt{2}v_{PQ}}T_{\tilde{a}\tilde{Z}h}^i,\\
\Lambda_{\tilde{a}\tilde{Z}H}^i&=&X_{i0}^H+X_{0i}^H-\frac{c_H\mu}{\sqrt{2}v_{PQ}}T_{\tilde{a}\tilde{Z}H}^i,\\
\Lambda_{\tilde{a}\tilde{Z}A}^i&=&X_{i0}^A+X_{0i}^A-\frac{c_H\mu}{\sqrt{2}v_{PQ}}T_{\tilde{a}\tilde{Z}A}^i,
\end{eqnarray}
where $X_{i0}^{h,H}$ is given in Eq.~(\ref{eq:Xhij}), $X^A_{ij}$ is given by
\begin{equation}
X^A_{ij}=\frac12(-1)^{\theta_i+\theta_j}\left(v_2^{(i)}\sin\beta-v_1^{(i)}\cos\beta\right)
\left(gv_3^{(j)}-g'v_4^{(j)}\right),
\end{equation}
and $T^i_{\tilde{a}\tilde{Z}\phi}$ are given by
\begin{eqnarray}
T^{i}_{\tilde{a}\tilde{Z}h}&=&(-1)^{\theta_i+\theta_{\tilde{a}}}\left(v_1^{(i)}\sin\alpha+v_2^{(i)}\cos\alpha\right),\\
T^{i}_{\tilde{a}\tilde{Z}H}&=&(-1)^{\theta_i+\theta_{\tilde{a}}}\left(v_1^{(i)}\cos\alpha-v_2^{(i)}\sin\alpha\right),\\
T^{i}_{\tilde{a}\tilde{Z}G^0}&=&(-1)^{\theta_i+\theta_{\tilde{a}}+1}\left(v_1^{(i)}\cos\beta-v_2^{(i)}\sin\beta\right),\\
T^{i}_{\tilde{a}\tilde{Z}A}&=&(-1)^{\theta_i+\theta_{\tilde{a}}+1}\left(-v_1^{(i)}\sin\beta+v_2^{(i)}\cos\beta\right).
\end{eqnarray}
$v_0^{(i)}$ and $v_i^{(0)}$ are axino-neutalino mixing components, which are given by
\begin{eqnarray}
v_0^{(0)}&=&1,\\
v_0^{(i)}&=&-\frac{c_H\mu v}{v_{PQ}}\frac{v_1^{(i)}\cos\beta+v_2^{(i)}\sin\beta}{m_{\tilde{a}}-m_{\widetilde{Z}_i}(-1)^{\theta_i}},\\
v_i^{(0)}&=&
\sum_{j=1}^4\frac{c_H\mu v}{v_{PQ}}\frac{v_i^{(j)}\left( v_1^{(j)}\cos\beta+ v_2^{(j)}\sin\beta\right)}{m_{\tilde{a}}-m_{\widetilde{Z}_j}(-1)^{\theta_j}},
\end{eqnarray}
for $i=1,\cdots,4$.

\item $\ta\to\widetilde{W}_i^{\pm}H^{\mp}$

\begin{eqnarray}
\Gamma(\tilde{a}\to\widetilde{W}_i^-H^+)&=&
\Gamma(\tilde{a}\to\widetilde{W}_i^+H^-)\nonumber\\
&=&\frac{1}{16\pi}m_{\tilde{a}}\lambda^{1/2}\left(1,\frac{m_{\widetilde{W}_i}^2}{m_{\tilde{a}}^2},\frac{m_{H^+}^2}{m_{\tilde{a}}^2}\right)\\
&&\times\left[
\left(a_i^2+b_i^2\right)\left(1+\frac{m_{\widetilde{W}_i}^2}{m_{\tilde{a}}^2}-\frac{m_{H^+}^2}{m_{\tilde{a}}^2}\right)
+2\left(a_i^2-b_i^2\right)\frac{m_{\widetilde{W}_i}}{m_{\tilde{a}}}
\right].
\label{dec_aWH}
\end{eqnarray}
The couplings are given by
\begin{eqnarray}
a_1&=&\frac12\left\{
(-1)^{\theta_{\widetilde{W}_1}}\cos\beta\Lambda_2^{(0)}
-(-1)^{\theta_{\tilde{a}}}\sin\beta\Lambda_4^{(0)}
\right\},\\
b_1&=&\frac12\left\{
(-1)^{\theta_{\widetilde{W}_1}}\cos\beta\Lambda_2^{(0)}
+(-1)^{\theta_{\tilde{a}}}\sin\beta\Lambda_4^{(0)}
\right\},\\
a_2&=&\frac12\left\{
(-1)^{\theta_{\widetilde{W}_2}}\theta_y\cos\beta\Lambda_1^{(0)}
-(-1)^{\theta_{\tilde{a}}}\theta_x\sin\beta\Lambda_3^{(0)}
\right\},\\
b_2&=&\frac12\left\{
(-1)^{\theta_{\widetilde{W}_2}}\theta_y\cos\beta\Lambda_1^{(0)}
+(-1)^{\theta_{\tilde{a}}}\theta_x\sin\beta\Lambda_3^{(0)}
\right\},
\end{eqnarray}
where
\begin{eqnarray}
\Lambda_1^{(0)}&=&A_1^{(0)}+\frac{c_H\mu}{v_{PQ}}(-1)^{\theta_{\tilde{a}}}\tan\beta\sin\gamma_R,\\
\Lambda_2^{(0)}&=&A_2^{(0)}-\frac{c_H\mu}{v_{PQ}}(-1)^{\theta_{\tilde{a}}}\tan\beta\cos\gamma_R,\\
\Lambda_3^{(0)}&=&A_3^{(0)}-\frac{c_H\mu}{v_{PQ}}(-1)^{\theta_{\tilde{a}}}\cot\beta\sin\gamma_L,\\
\Lambda_4^{(0)}&=&A_4^{(0)}+\frac{c_H\mu}{v_{PQ}}(-1)^{\theta_{\tilde{a}}}\cot\beta\cos\gamma_L,
\end{eqnarray}
and
\begin{eqnarray}
A_1^{(0)}&=&-\frac{1}{\sqrt{2}}
\left(gv_3^{(0)}+g'v_4^{(0)}\right)\sin\gamma_R-gv_1^{(0)}\cos\gamma_R,\\
A_2^{(0)}&=&\frac{1}{\sqrt{2}}
\left(gv_3^{(0)}+g'v_4^{(0)}\right)\cos\gamma_R-gv_1^{(0)}\sin\gamma_R,\\
A_3^{(0)}&=&-\frac{1}{\sqrt{2}}
\left(gv_3^{(0)}+g'v_4^{(0)}\right)\sin\gamma_L+gv_2^{(0)}\cos\gamma_L,\\
A_4^{(0)}&=&\frac{1}{\sqrt{2}}
\left(gv_3^{(0)}+g'v_4^{(0)}\right)\cos\gamma_L+gv_2^{(0)}\sin\gamma_L.
\end{eqnarray}
$\theta_{\ta}$ is 1(0) if axino mass term is negative (positive).

\item $\ta\to \widetilde{Z}_iZ$

\begin{eqnarray}
\Gamma(\tilde{a}\to\widetilde{Z}_iZ)&=&
\frac{1}{4\pi}|W_{i0}|^2m_{\tilde{a}}\lambda^{1/2}\left(1,\frac{m_{\widetilde{Z}_i}^2}{m_{\tilde{a}}^2},\frac{M_{Z}^2}{m_{\tilde{a}}^2}\right)\nonumber\\
&&\times\left[
\left(1+\frac{m_{\widetilde{Z}_i}^2}{m_{\tilde{a}}^2}-2\frac{M_{Z}^2}{m_{\tilde{a}}^2}\right)\right.\nonumber\\
&&\left.+\left(\frac{m_{\tilde{a}}^2}{M_{Z}^2}\right)
\left(1-\frac{m_{\widetilde{Z}_i}^2}{m_{\tilde{a}}^2}\right)^2
+6(-1)^{\theta_i+\theta_{\tilde{a}}}\frac{m_{\widetilde{Z}_i}}{m_{\tilde{a}}}
\right],
\label{dec_aZZ}
\end{eqnarray}
where
\begin{equation}
W_{i0}=\frac14\sqrt{g^2+g'^2}(-i)^{\theta_i}(i)^{\theta_{\tilde{a}}}
\left(v_1^{(i)}v_1^{(0)}-v_2^{(i)}v_2^{(0)}\right).
\end{equation}

\item $\ta\to \widetilde{W}_i^{\pm}W^{\mp}$

\begin{eqnarray}
\Gamma(\tilde{a}\to\widetilde{W}_i^-W^+)&=&
\Gamma(\tilde{a}\to\widetilde{W}_i^+W^-)\nonumber\\
&=&\frac{g^2}{16\pi}m_{\tilde{a}}\lambda^{1/2}\left(1,\frac{m_{\widetilde{W}_i}^2}{m_{\tilde{a}}^2},\frac{M_{W}^2}{m_{\tilde{a}}^2}\right)\nonumber\\
&&\times\Biggl[
\left({X_i^0}^2+{Y_i^0}^2\right)
\left\{
\left(1+\frac{m_{\widetilde{W}_i}^2}{m_{\tilde{a}}^2}-2\frac{M_{W}^2}{m_{\tilde{a}}^2}\right)
\right.\nonumber\\
&&\left.+\left(\frac{m_{\tilde{a}}^2}{M_{W}^2}\right)
\left(1-\frac{m_{\widetilde{W}_i}^2}{m_{\tilde{a}}^2}\right)^2
\right\}-6\left({X_i^0}^2-{Y_i^0}^2\right)\frac{m_{\widetilde{W}_i}}{m_{\tilde{a}}}
\Biggr].
\label{dec_aWW}
\end{eqnarray}
The couplings are given by
\begin{eqnarray}
X_1^0&=&\frac12\left[
(-1)^{\theta_{\widetilde{W}_1}+\theta_{\tilde{a}}}
\left(\frac{\cos\gamma_R}{\sqrt{2}}v_1^{(0)}+\sin\gamma_Rv_3^{(0)}\right)\right.\nonumber\\
&&\left.
-\frac{\cos\gamma_L}{\sqrt{2}}v_2^{(0)}+\sin\gamma_Lv_3^{(0)}
\right],\\
X_2^0&=&\frac12\left[
(-1)^{\theta_{\widetilde{W}_2}+\theta_{\tilde{a}}}\theta_y
\left(\frac{-\sin\gamma_R}{\sqrt{2}}v_1^{(0)}+\cos\gamma_Rv_3^{(0)}\right)\right.\nonumber\\
&&\left.+\theta_x\left(\frac{\sin\gamma_L}{\sqrt{2}}v_2^{(0)}+\cos\gamma_Lv_3^{(0)}\right)
\right],\\
Y_1^0&=&\frac12\left[
-(-1)^{\theta_{\widetilde{W}_1}+\theta_{\tilde{a}}}
\left(\frac{\cos\gamma_R}{\sqrt{2}}v_1^{(0)}+\sin\gamma_Rv_3^{(0)}\right)\right.\nonumber\\
&&\left.
-\frac{\cos\gamma_L}{\sqrt{2}}v_2^{(0)}+\sin\gamma_Lv_3^{(0)}
\right],\\
Y_2^0&=&\frac12\left[
-(-1)^{\theta_{\widetilde{W}_2}+\theta_{\tilde{a}}}\theta_y
\left(\frac{-\sin\gamma_R}{\sqrt{2}}v_1^{(0)}+\cos\gamma_Rv_3^{(0)}\right)\right.\nonumber\\
&&\left.
+\theta_x\left(\frac{\sin\gamma_L}{\sqrt{2}}v_2^{(0)}+\cos\gamma_Lv_3^{(0)}\right)
\right].
\end{eqnarray}

\item $\ta\to f\tilde{f}_k$

\begin{eqnarray}
\Gamma(\tilde{a}\to f \tilde{f}_k)&=&\frac{N_c m_{\tilde{a}}}{16\pi}\lambda^{1/2}\left(1,\frac{m_{\tilde{f}_k}^2}{m_{\tilde{a}}^2},\frac{m_f^2}{m_{\tilde{a}}^2}\right)\nonumber\\
&&\times
\left[
|a_f^k|^2\left\{\left(1+\frac{m_f}{m_{\tilde{a}}}\right)^2-\frac{m_{\tilde{f}_k}^2}{m_{\tilde{a}}^2}\right\}
\right.\nonumber\\
&&\left.+|b_f^k|^2\left\{\left(1-\frac{m_f}{m_{\tilde{a}}}\right)^2-\frac{m_{\tilde{f}_k}^2}{m_{\tilde{a}}^2}\right\}
\right],
\end{eqnarray}
where
\begin{eqnarray}
a_u^1&=&\frac12\left[
\left\{iA_{\tilde{a}}^u-(i)^{\theta_{\tilde{a}}}f_uv_1^{(0)}\right\}\cos\theta_u
-\left\{iB_{\tilde{a}}^u-(-i)^{\theta_{\tilde{a}}}f_uv_1^{(0)}\right\}\sin\theta_u
\right],\\
b_u^1&=&\frac12\left[
\left\{-iA_{\tilde{a}}^u-(i)^{\theta_{\tilde{a}}}f_uv_1^{(0)}\right\}\cos\theta_u
-\left\{iB_{\tilde{a}}^u+(-i)^{\theta_{\tilde{a}}}f_uv_1^{(0)}\right\}\sin\theta_u
\right],\\
a_d^1&=&\frac12\left[
\left\{iA_{\tilde{a}}^d-(i)^{\theta_{\tilde{a}}}f_dv_2^{(0)}\right\}\cos\theta_d
-\left\{iB_{\tilde{a}}^d-(-i)^{\theta_{\tilde{a}}}f_dv_2^{(0)}\right\}\sin\theta_d
\right],\\
b_d^1&=&\frac12\left[
\left\{-iA_{\tilde{a}}^d-(i)^{\theta_{\tilde{a}}}f_dv_2^{(0)}\right\}\cos\theta_d
-\left\{iB_{\tilde{a}}^d+(-i)^{\theta_{\tilde{a}}}f_dv_2^{(0)}\right\}\sin\theta_d
\right],
\end{eqnarray}
\begin{eqnarray}
a_u^2&=&\frac12\left[
\left\{iA_{\tilde{a}}^u-(i)^{\theta_{\tilde{a}}}f_uv_1^{(0)}\right\}\sin\theta_u
+\left\{iB_{\tilde{a}}^u-(-i)^{\theta_{\tilde{a}}}f_uv_1^{(0)}\right\}\cos\theta_u
\right],\\
b_u^2&=&\frac12\left[
\left\{-iA_{\tilde{a}}^u-(i)^{\theta_{\tilde{a}}}f_uv_1^{(0)}\right\}\sin\theta_u
+\left\{iB_{\tilde{a}}^u+(-i)^{\theta_{\tilde{a}}}f_uv_1^{(0)}\right\}\cos\theta_u
\right],\\
a_d^2&=&\frac12\left[
\left\{iA_{\tilde{a}}^d-(i)^{\theta_{\tilde{a}}}f_dv_2^{(0)}\right\}\sin\theta_d
+\left\{iB_{\tilde{a}}^d-(-i)^{\theta_{\tilde{a}}}f_dv_2^{(0)}\right\}\cos\theta_d
\right],\\
b_d^2&=&\frac12\left[
\left\{-iA_{\tilde{a}}^d-(i)^{\theta_{\tilde{a}}}f_dv_2^{(0)}\right\}\sin\theta_d
+\left\{iB_{\tilde{a}}^d+(-i)^{\theta_{\tilde{a}}}f_dv_2^{(0)}\right\}\cos\theta_d
\right],
\end{eqnarray}
with
\begin{eqnarray}
A_{\tilde{a}}^u&=&\frac{(-i)^{\theta_{\tilde{a}}-1}}{\sqrt{2}}
\left[gv_3^{(0)}+\frac{g'}{3}v_4^{(0)}\right],\\
A_{\tilde{a}}^d&=&\frac{(-i)^{\theta_{\tilde{a}}-1}}{\sqrt{2}}
\left[-gv_3^{(0)}+\frac{g'}{3}v_4^{(0)}\right],\\
B_{\tilde{a}}^u&=&\frac{4}{3\sqrt{2}}g'(i)^{\theta_{\tilde{a}}-1}v_4^{(0)},\\
B_{\tilde{a}}^d&=&-\frac{2}{3\sqrt{2}}g'(i)^{\theta_{\tilde{a}}-1}v_4^{(0)}.
\end{eqnarray}
Here $f_{u,d}$ is the Yukawa coupling constant.

\end{itemize}

\end{document}